\documentclass[11pt,a4paper]{article}	

\usepackage{color} 
\usepackage{amsmath}
\usepackage{amsfonts}
\usepackage{amssymb}
\usepackage{float}
\usepackage{caption}
\usepackage{graphicx}
\usepackage{slashed}            
\usepackage{subfig}             
\usepackage{bbm}                
\usepackage{xspace}				
\usepackage{collref}				
\usepackage{framed}				
\usepackage{placeins}
\usepackage{jheppub}				

\usepackage{tikz}
\usetikzlibrary{arrows,shapes}
\usetikzlibrary{trees}
\usetikzlibrary{matrix,arrows} 				
\usetikzlibrary{positioning}				
\usetikzlibrary{calc,through}				
\usetikzlibrary{decorations.pathreplacing}  
\usepackage{pgffor}							

\usetikzlibrary{decorations.pathmorphing}	
\usetikzlibrary{decorations.markings}
\usetikzlibrary{snakes}

\tikzset{
    vector/.style={decorate, decoration={snake}, draw},
	provector/.style={decorate, decoration={snake,amplitude=2.5pt}, draw},
	antivector/.style={decorate, decoration={snake,amplitude=-2.5pt}, draw},
    fermion/.style={draw, postaction={decorate},
        decoration={markings,mark=at position .55 with {\arrow[draw]{>}}}},
    fermionbar/.style={draw, postaction={decorate},
        decoration={markings,mark=at position .55 with {\arrow[draw=black]{<}}}},
    fermionnoarrow/.style={draw},
    gluon/.style={decorate, draw,decoration={coil,amplitude=4pt, segment length=6pt}, line width=1},
    scalar/.style={dashed,draw, postaction={decorate},
        decoration={markings,mark=at position .55 with {\arrow[draw]{>}}}},
    scalarbar/.style={dashed,draw, postaction={decorate},
        decoration={markings,mark=at position .55 with {\arrow[draw]{<}}}},
    scalarnoarrow/.style={dash pattern = on 6 pt off 3 pt,draw},
    electron/.style={draw, postaction={decorate},
        decoration={markings,mark=at position .55 with {\arrow[draw]{>}}}},
	bigvector/.style={decorate, decoration={snake,amplitude=4pt}, draw},
	vectorscalar/.style={loosely dotted,draw, postaction={decorate}},
}

\renewcommand{\tilde}{\widetilde} 

\newcommand{\tF}[5]{\tilde{F}^{#2 #4 #5}_{#1,#3}}
\newcommand{\tDF}[5]{\tilde{D}_{#1} \tilde{F}^{#2 #4 #5}_{#1,#3}}

\usepackage{hyperref}   

\title{\boldmath The birds and the $B$s in RS: the $b\to s\gamma$ penguin in a warped extra dimension}

\author[a]{Monika Blanke,}
\author[a]{Bibhushan Shakya,}
\author[a]{Philip Tanedo,}
\author[a]{and Yuhsin Tsai}
\affiliation[a]{Department of Physics, LEPP, Cornell University,\\
142 Sciences Drive, Ithaca, NY 14853, USA}
\emailAdd{mb744@cornell.edu}
\emailAdd{bs475@cornell.edu}
\emailAdd{pt267@cornell.edu}
\emailAdd{yt237@cornell.edu}

\abstract{
We calculate contributions to the photon and gluon magnetic dipole operators that mediate $b\to s\gamma$ and $b\to d\gamma$ transitions in the Randall-Sundrum model of a warped extra dimension with anarchic bulk fermions and a brane localized Higgs. 
Unlike the Standard Model, there are large contributions to the left-handed $b$ quark decays, parameterized by the Wilson coefficient $C'_7$, due to the pattern of bulk fermion localization, and sizable contributions from the gluonic penguins, $C_8^{(\prime)}$, through renormalization group mixing. 
Further, unlike the Randall-Sundrum result for $\mu\to e\gamma$, the unprimed Wilson coefficients receive non-negligible contributions from the misalignment of the bulk fermion spectrum with the Standard Model flavor sector. 
We compare the size of effects and the constraints imposed by the branching ratios $\text{Br}(B\to X_s\gamma)$ and $\langle \text{Br}(B\to X_d\gamma) \rangle$ within the minimal and the custodial model.
Within the custodial framework, we study the effect on a number of benchmark observables and find that $\text{Br}(B\to X_s\mu^+\mu^-)$ and the forward-backward asymmetry in $B\to K^*\mu^+\mu^-$ remain close to their Standard Model predictions. On the other hand, there can be large enhancements of the time-dependent CP asymmetry in $B\to K^*\gamma$ and the transverse asymmetry~$A_T^{(2)}$.
}

\arxivnumber{1203.6650}
 
\begin{document}
\maketitle

\section{Introduction}

The Randall-Sundrum (RS) scenario of a warped extra dimension provides an elegant solution to the hierarchy problem \cite{Randall:1999ee, Pomarol:1999ad, Gherghetta:2000qt, Davoudiasl:1999tf, Grossman:1999ra, Chang:1999nh} and a way to understand strongly coupled dynamics through the AdS/CFT correspondence \cite{Maldacena:1997re, ArkaniHamed:2000ds, Rattazzi:2000hs}\nocollect{Pomarol:1999ad}. For reviews see \cite{Csaki:2004ay, Csaki:2005vy, Gherghetta:2010cj, Sundrum:2011ic}. One of the promising phenomenological features to come out of this framework is an explanation of the Standard Model (SM) flavor structure through the split-fermion scenario \cite{Gherghetta:2000qt, Grossman:1999ra, ArkaniHamed:1999dc, Grossman:2002pb}. In these models the Yukawa matrices are anarchic and the spectrum of fermion masses is generated by the exponential suppression of zero mode wavefunctions with a brane-localized Higgs~\cite{Agashe:2004cp}.
This also automatically generates hierarchical mixing angles~\cite{Agashe:2004cp, Huber:2003tu,Kitano:ij} and suppresses many tree-level flavor-changing neutral currents (FCNCs) through the RS-GIM mechanism \cite{Agashe:2004cp}.
In order to protect against large contributions to the $T$ parameter coming from bulk gauge fields, one may introduce a gauged custodial symmetry \cite{Agashe:2003zs} that is broken on the boundaries; a straightforward discrete extension of such a symmetry also protects against corrections to the $Zb\bar{b}$ vertex \cite{Agashe:2006at,Contino:2006qr} and flavor changing couplings of the $Z$ boson to left-handed down-type quarks \cite{Blanke:2008yr,Albrecht:2009xr}.

These flavor protection mechanisms are not always sufficient to completely protect RS models from stringent experimental flavor constraints. 
In the quark sector, the tree-level exchange of Kaluza-Klein (KK) gluons and neutral electroweak gauge bosons contributes to meson-antimeson mixing and induces left-right operators. These operators are not present in the SM and receive a significant enhancement through QCD effects due to their large anomalous dimension. In the kaon system they are also chirally enhanced by a factor of $m_K^2/m_s^2$.
These contributions lead to new  CP violating effects in the kaon system, namely the well-measured observable $\varepsilon_K$, and result in generic bounds of ${\cal O}(10-20$ TeV) for the KK gluon mass~\cite{Albrecht:2009xr,Csaki:2008zd,Blanke:2008zb,Buras:2009ka,Duling:2009pj,Bauer:2009cf,Casagrande:2008hr}. 
To reduce this bound, one must introduce additional structure such as horizontal symmetries~\cite{Santiago:2008vq, Csaki:2008eh}, flavor alignment~\cite{Fitzpatrick:2008zza, Csaki:2009wc}, or an extended strong sector~\cite{Bauer:2011ah}. Alternately, one may promote the Higgs to a bulk field~\cite{Agashe:2008uz} to localize the fermion zero modes closer to the UV brane.
%

Additional constraints on the RS flavor sector come from loop-induced dipole operators through penguin diagrams. The first estimates for these operators were performed in~\cite{Agashe:2004cp,Huber:2003tu,Agashe:2004ay,Agashe:2006iy} assuming UV sensitivity at all loops within the 5D effective theory and a calculation within the two-site approach was performed in \cite{Agashe:2008uz}. In \cite{Agashe:2004cp} the bound $M_\text{KK} > \mathcal{O}(10\,\text{TeV})$ was derived from the constraint on the neutron electric dipole moment. 
%
%
The RS dipole contributions lead to dangerously large effects in direct CP violation in the $K\to\pi\pi$ decays measured by the ratio $\varepsilon'/\varepsilon$ \cite{Gedalia:2009ws}. 
Combining the bound from the latter ratio with the $\varepsilon_K$ constraint leads to a lower bound on the KK scale independent of the strength of the 5D Yukawa.
More recently it was shown that even for the brane Higgs scenario the one-loop induced magnetic penguin diagrams are finite in RS and can be calculated effectively in a manifestly 5D formalism \cite{Csaki:2010aj}. The lepton flavor violating penguin $\mu\to e\gamma$ sets bounds on the KK and anarchic Yukawa scales that are complementary to tree-level processes, so the tension between these bounds quantifies the degree of tuning required in the 5D Yukawa matrix \cite{Agashe:2006iy}.

In this paper we examine the calculation and phenomenological observables of the quark sector processes $b\to q\gamma$ ($q=s,d$) in the RS framework with a brane-localized Higgs field using the mixed position--momentum space formalism. These processes differ from their leptonic analogs for various reasons beyond the spectrum and diagrams involved. 
Firstly, while the branching ratio of $\mu\to e\gamma$ is only bounded from above, the branching ratios for $B\to X_s\gamma$ and, to a lesser extent, $B\to X_d\gamma$ are well-measured and in good agreement with the SM. Secondly, theoretical predictions are more involved due to the renormalization group (RG) evolution from the KK scale to the $B$ meson scale and hadronic effects at the latter scale.
The RG running over this large range of energy scales introduces a sizable mixing between the various effective operators, so that one must also include the effects of the magnetic gluon penguin $C^{(\prime)}_8$ in addition to the magnetic photon penguin $C^{(\prime)}_7$.

After reviewing the flavor structure of RS models in Section~\ref{sec:flavor:in:RS}, we calculate the $C_7^{(\prime)}$ and $C_8^{(\prime)}$ Wilson coefficients of the quark dipole operators in Section~\ref{sec:penguins}. 
We provide explicit formulae for the dominant RS contributions to the Wilson coefficients at the KK scale in both the minimal and custodial models and analyze the size of these contributions. In Sections~\ref{sec:BXsdgam} and \ref{sec:bsmumu}, we subsequently perform the RG evolution down to the $B$ meson scale and obtain predictions for the branching ratios $\text{Br}(B\to X_{s,d}\gamma)$.

Finally, in Section~\ref{sec:num}, we investigate the phenomenological implications on a number of benchmark observables related to the photon and gluon penguin operators. We first show that these operators give non-negligible constraints for both minimal and custodial models. We then restrict our attention to realistic models with a bulk custodial symmetry $SU(2)_L\times SU(2)_R\times U(1)_X\times P_{LR}$ and consider the effect of benchmark observables on points in parameter space that pass tree-level constraints as evaluated in \cite{Blanke:2008zb}.
Rather than performing a detailed analysis of all observables provided by the $B\to X_s\gamma$, $B\to K^*\gamma$, $B\to X_s\mu^+\mu^-$ and $B\to K^*\mu^+\mu^-$ decay modes, we focus on a number of benchmark observables in order to illustrate the pattern of effects and leave a more detailed analysis for future work.
Specifically we study:
\begin{itemize}
\item
The branching ratio $\text{Br}(B\to X_s\gamma)$ and the CP averaged branching ratio $\langle \text{Br}(B\to X_d\gamma)\rangle$ which we impose as constraints on our parameter scan. 
\item
The branching ratio $\text{Br}(B\to X_s\mu^+\mu^-)$ and the forward backward asymmetry $A_\text{FB}$ in $B\to K^*\mu^+\mu^-$. Stringent data that are in good agreement with the SM exist for both observables, placing strong bounds on various new physics (NP) scenarios. The custodial RS model naturally predicts small effects in these observables since they are rather insensitive to NP contributions to the primed magnetic Wilson coefficients.
\item
The time-dependent CP asymmetry $S_{K^*\gamma}$ in $B\to K^*\gamma$ and the transverse asymmetry $A_T^{(2)}$ in $B\to K^*\mu^+\mu^-$, evaluated in the region of low dimuon invariant mass $1\,\text{GeV}^2<q^2<6\,\text{GeV}^2$.
\end{itemize}

Since the RS contributions generally exhibit the hierarchy $\Delta C'_7\gg \Delta C_7$ \cite{Agashe:2004cp, Agashe:2004ay} the latter observables are particularly suited to look for RS contributions. CP asymmetries in radiative $B$ decays were already suggested in \cite{Agashe:2004cp,Agashe:2004ay} as good probes to look for RS effects. We quantify the possible size of effects and study the possible RS contributions to the various observables in a correlated manner. We also included the transverse asymmetry $A_T^{(2)}$, which has not been considered in the context of RS models before.

\section{Flavor in Randall-Sundrum models}
\label{sec:flavor:in:RS}

We summarize here the relevant aspects of flavor physics and the RS scenario. For a review of the general framework see e.g.~\cite{Csaki:2005vy, Gherghetta:2010cj, Sundrum:2011ic,Bauer:2009cf,Albrecht:2009xr}. 
We consider a 5D warped interval $z\in[R,R']$ with an infrared (IR) brane at $z=R' \sim (\text{TeV})^{-1}$ and an ultraviolet (UV) brane at $z=R \sim M_\text{Pl}$, the AdS curvature scale. In conformal coordinates the metric is
\begin{align}\label{eq:metric}
    ds^2&=\left(\frac{R}{z}\right)^2 (dx_\mu dx_\nu \eta^{\mu\nu} -dz^2).
\end{align}
One may recover the classic RS conventions with the identifications $z = R\exp(ky)$ and $k=1/R$, $k\exp{(-kL)}=1/R'$.
Fermions are Dirac fields that propagate in the bulk and can be written in terms of left- and right-handed Weyl spinors $\chi$ and $\bar\psi$ via
\begin{align}
    \Psi(x,z) =
        \begin{pmatrix}
        \chi(x,z) \\ \bar\psi(x,z)
        \end{pmatrix}.
\end{align}
In order to obtain a spectrum with chiral zero modes, fermions must have chiral (orbifold) boundary conditions,
\begin{align}
    \psi_L(x^\mu,R)=\psi_L(x^\mu,R') = 0 \quad\quad\quad\text{ and }\quad\quad\quad \chi_R(x^\mu,R) =\chi_R(x^\mu,R') = 0,
\end{align}
where the subscripts $L$ and $R$ denote the SU$(2)_\text{L}$ doublet ($L$) and singlet ($R$) representations, i.e.\ the chirality of the zero mode (SM fermion).
The localization of the normalized zero mode profile is controlled by the dimensionless parameter $c$, 
\begin{align}
    \chi^{(0)}_{c} (x,z) &=\frac{1}{\sqrt{R'}} \left(\frac{z}{R}\right)^2 \left(\frac{z}{R'}\right)^{-c} f_c \;\chi^{(0)}_c(x)
    \quad\quad\quad \text{and }\quad\quad\quad  \psi_c^{(0)} (x,z) = \chi_{-c}^{(0)} (x,z), \label{eq:fermion:profile}
\end{align}
where $c/R$ is the fermion bulk mass.
Here we have defined the RS flavor function characterizing the fermion profile on the IR brane,
\begin{align}
f_c &=\sqrt{\frac{1-2c}{1-(R/R')^{1-2c}}}.\label{eq:flavor:function}
\end{align}

We assume that the Higgs is localized on the IR brane.
The Yukawa coupling is 
\begin{align}
 S_{\text{Yuk}} =  \int d^4 x\; \left(\frac{R}{R'}\right)^4  
\left[ - \frac{1}{\sqrt{2}}
\left(
\bar Q_i \cdot \tilde H \,R\,Y_{u,ij} U_j 
+ \bar Q_i \cdot H \,R\,Y_{d,ij} D_j
   + \text{h.c.}
\right)\right]
 \label{eq:Yukawa:action}
\end{align}
where $Y_{ij}$ are dimensionless 3$\times$3  matrices such that $(Y_5)_{ij}=RY_{ij}$ is the dimensionful parameter appearing in the 5D Lagrangian with $Y$ assumed to be a random `anarchic' matrix with average elements of order $Y_*$. After including warp factors and canonically normalizing fields, the effective 4D Yukawa and zero mode mass matrices are 
\begin{align}
y^{\text{SM}}_{ij} &= f_{c_{L_i}} Y_{ij} f_{-c_{R_j}} \quad\quad\quad\quad\quad\quad m_{ij} = \frac{v}{\sqrt{2}}y^{\text{SM}}_{ij},
\label{eq:RS:anarchy:zero:mode:Yukawa}
\end{align}
so that the fermion mass hierarchy is set by the $f_1 \ll f_2 \ll f_3$ structure for both left- and right-handed zero modes. At the same time, the hierarchical pattern of the CKM matrix is also generated naturally. In other words, the choice of $c$ for each fermion family introduces additional flavor structure into the theory that generates the zero mode spectrum while allowing the fundamental Yukawa parameters to be anarchic. 

In this document we work in the gauge basis where the bulk mass matrices and the interactions of the neutral gauge bosons are flavor diagonal but not flavor universal. The Yukawa couplings are non-diagonal in this basis and cause the resulting fermion mass matrices to be non-diagonal. Since these off-diagonal entries are governed by the small parameter $vR'$, we will treat them as a perturbative correction in the mass insertion approximation.

Realistic RS models typically require a mechanism to suppress generically large contributions to the Peskin-Takeuchi $T$ parameter and the $Zb\bar b$ coupling; a common technique is to extend the bulk gauge symmetry to \cite{Agashe:2003zs,Csaki:2003zu,Agashe:2004rs,Agashe:2006at,Contino:2006qr,Carena:2006bn,Cacciapaglia:2006gp}
\begin{equation}
SU(3)_c\times SU(2)_L\times SU(2)_R\times U(1)_X\times P_{LR}
.
\end{equation}
Here $P_{LR}$ is a discrete symmetry exchanging the $SU(2)_L$ and $SU(2)_R$ factors; in order to protect the left-handed $Zb\bar b$ coupling from anomalously large corrections, the left-handed down type quarks have to be eigenstates under $P_{LR}$. This in turn requires enlarged fermion representations with respect to the minimal model. 
Specifically the quark representations containing the SM zero modes are $(i=1,2,3)$:
\begin{eqnarray}\label{eq:xi1L}
\xi^i_{1L}&=&\begin{pmatrix}\chi^{u_i}_{L}(-+)_{5/3} && q_L^{u_i}(++)_{2/3} \\
\chi^{d_i}_{L}(-+)_{2/3} && q_L^{d_i}(++)_{-1/3}\end{pmatrix}_{2/3}\,,\\
\xi^i_{2R} &=& u^i_R (++)_{2/3}\,,\label{eq:xi2R}\\
\xi^i_{3R} &=& T^i_{3R} \oplus T^i_{4R} = \begin{pmatrix}
\psi^{\prime i}_R(-+)_{5/3} \\ U^{\prime i}_R (-+)_{2/3} \\ D^{\prime i}_R (-+)_{-1/3} \end{pmatrix}_{2/3} \oplus 
\begin{pmatrix}
\psi^{\prime\prime i}_R(-+)_{5/3} \\ U^{\prime\prime i}_R (-+)_{2/3} \\ D^{ i}_R (++)_{-1/3} \end{pmatrix}_{2/3}\,.\label{eq:xi3R}
\end{eqnarray}
Here $\xi^i_{1L}$ is an $SU(2)_L \times SU(2)_R$ bidoublet,  $\xi^i_{2R}$ is singlet under both $SU(2)$s, and $T^i_{3R}$ and $T^i_{4R}$ are triplets under $SU(2)_L$ and $SU(2)_R$ respectively, with all of them carrying $U(1)_X$ charge $+2/3$.
The corresponding states of opposite chirality are obtained by reversing the boundary conditions.

As we will see later, while the additional gauge bosons present in the custodial model do not have a significant impact on the $b\to q\gamma$ and $b\to qg$ ($q=d,s$) amplitudes, the additional fermion modes contribute and generally enhance the effect.

\section{\boldmath Calculation of the $b\to q\gamma$ Penguin in RS}\label{sec:penguins}

We now calculate the RS contributions to the $b\to q\gamma$ and $b\to qg$ ($q=d,s$) decays. These contributions are calculated at the KK scale $M_\text{KK}\sim 1/R'$; in subsequent sections we will relate these to renormalization group (RG) evolved coefficients and observables at the low scale $\sim m_b$. 

We only evaluate the dominant diagrams, working in Feynman gauge and the mass insertion approximation, where the expansion parameter is $vR'/\sqrt{2} \sim \mathcal O(0.1)$.
We have checked explicitly that the diagrams presented here dominate those that were neglected by at least an order of magnitude; a more detailed calculation is beyond the scope of this work and, in our opinion, premature before the discovery of RS KK modes.
We refer to \cite{Csaki:2010aj} for details of the 5D calculation, Feynman rules, and guidelines for estimating the dominant diagrams. For additional notation and conventions, especially with respect to the custodially protected model, see \cite{Albrecht:2009xr}. See Appendix~\ref{sec:theory:uncertainty} for comments on theory uncertainties.

\subsection[Effective Hamiltonian for $b\to q\gamma$ transitions]{\boldmath Effective Hamiltonian for $b\to q\gamma$ transitions} 

The $b\to q\gamma$ $(q=d,s)$ transitions are most conveniently described by an effective Hamiltonian in the operator product expansion, see e.\,g.\ \cite{Buchalla:1995vs} for a review. The dipole terms most sensitive to new physics are
\begin{eqnarray}\label{eq:MasterHamiltonian}
\mathcal{H}_\text{eff} = - \frac{G_F}{\sqrt{2}}V_{tq}^*V_{tb}
\Big[
C_7(\mu)Q_{7}(\mu)+C'_{7}(\mu)Q'_{7}(\mu)+
C_8(\mu)Q_{8}(\mu)+C'_{8}(\mu)Q'_{8}(\mu)\Big]+\text{h.c.},\qquad
\end{eqnarray}
where we neglect terms proportional to $V_{uq}^* V_{ub}$.  The effective operators are 
\begin{align}
&Q_{7} = \frac{e}{4\pi^2}m_b (\bar q \sigma_{\mu\nu}P_R b)F^{\mu\nu} && Q'_{7} = \frac{e}{4\pi^2}m_b (\bar q \sigma_{\mu\nu}P_L b)F^{\mu\nu}\\
&Q_{8} = \frac{g_s}{4\pi^2}m_b (\bar q \sigma_{\mu\nu}T^a P_R b)G^{\mu\nu,a} && Q'_{8} = \frac{g_s}{4\pi^2}m_b (\bar q \sigma_{\mu\nu}T^a P_L b)G^{\mu\nu,a},
\end{align}
where $P_{L,R}=(1\mp \gamma^5)/2$. In this document we will focus on new contributions from the RS model to these operators. There are also contributions from non-dipole operators $Q_{1,\dots,6}$ and their chirality-flipped (primed) counterparts, but these are far less sensitive to NP and can be assumed to be equal to their SM contributions\footnote{The impact of flavor changing neutral gauge bosons on the operators $Q_{1,\dots,6}$ has recently been studied in \cite{Buras:2011zb}. Since the relevant contributions in RS are suppressed both by the KK scale and the RS GIM mechanism, the contributions are expected to be small and will be neglected in this paper.}.

At leading order in the SM, the primed Wilson coefficients $C'_{7,8}$ are suppressed by $m_s/m_b$ and therefore negligible, so the relevant Wilson coefficients at the scale $M_W$ are
\begin{equation}
C_7^{\text{SM}}(M_W) = -\frac{1}{2}D'_0(x_t)\,,\qquad C_8^{\text{SM}}(M_W) = -\frac{1}{2}E'_0(x_t),\label{eq:C7:and:C8:SM}
\end{equation}
where $x_t = m_t^2/M_W^2$, and $D'_0(x_t)\approx 0.37$ and $E'_0(x_t)\approx 0.19$ are loop functions given explicitly in (3.15--3.16) of \cite{Buras:1998raa}. In what follows we refer to the RS contributions to these  operators as $\Delta C_{7,8}^{(\prime)}$.


\subsection{Structure of the amplitude}
\label{sec:structure:of:amplitude}

In order to calculate the $b\to (s,d)\gamma$ and $b\to (s,d) g$ penguins, we work in a manifestly 5D framework. Unlike the 4D KK reduction, this procedure automatically incorporates the entire KK tower\footnote{
An alternate method of including the entire KK tower based on residue theorems was presented in \cite{Feng:2011pua}, though it obfuscates the physical intuition presented below.
}
at the cost of an expansion with respect to the Higgs-induced mass term ($\sim vR'$).

Using the on-shell condition for the photon, the general form of the left-to-right chirality $f_i^L(p)\to f_j^R(p') \gamma$ amplitude, $C_7$, in a 5D theory can be written as \cite{Csaki:2010aj,Lavoura:2003xp} 
\begin{align}\label{eq:MasterAmplitude}
\mathcal{M}_{L^i\to R^j}
=
\frac{ie}{16\pi^2}
\frac{vR'^2}{\sqrt{2}}
\displaystyle{\sum_{k,\ell}}
\left(\,a_{k\ell}Y^{\dagger}_{ik}Y_{k\ell}Y^{\dagger}_{\ell j}+b_{ij}Y_{ij}^\dag\right)
f_{Q_i}f_{D_{j}}
\bar u_{p'}^R\left[(p+p')^\mu -(m_b+m_q)\gamma^\mu \right]u_p^L\epsilon_\mu
\end{align}
where $\epsilon$ is the photon polarization. 
The chirality flipped amplitude is given by the conjugate of this result, $\mathcal{M}_{R^i\to L^j}=(\mathcal{M}_{L^j\to R^i})^\dagger$.
%
%
%
%
The expression for the gluon penguin is analogous with the appropriate substitutions. Using the fermion equations of motion, the term in the square brackets gives the required dipole structure $\sigma^{\mu\nu}F_{\mu\nu}$, so a simple way to identify the gauge-invariant contribution to the amplitude is to determine the coefficient of the $(p+p')^\mu$ term \cite{Lavoura:2003xp}. In \cite{Csaki:2010aj} this observation was used to show the manifest one-loop finiteness of these dipole transitions in 5D theories. 
Matching (\ref{eq:MasterAmplitude}) to the effective Hamiltonian (\ref{eq:MasterHamiltonian}) yields expressions for the RS contributions to the Wilson coefficients, $\Delta C$.

We refer to the coefficients $a_{k\ell}$ and $b_{ij}$ in (\ref{eq:MasterAmplitude}) as the \emph{anarchic} and the \emph{misalignment} contributions, respectively. They are products of couplings and dimensionless integrals whose flavor indices reflect the bulk mass dependence of internal propagators. 
%
Upon diagonalizing the SM fermion mass matrix, the anarchic term $a$ is not diagonalized and generally remains anarchic.
%
On the other hand, in the limit where the bulk masses are degenerate, the flavor structure of the $b$ term is aligned with the SM Yukawa matrices and thus contains no flavor-changing transitions in the mass basis \cite{Agashe:2004cp, Agashe:2008uz, Agashe:2006iy}. This alignment is pronounced for the first and second generation fermions because their bulk masses are nearly degenerate, but special care is required for the third generation quarks since these are localized towards the IR brane. 
The physical contribution of the $b$ coefficient comes from the robustness of off-diagonal elements of $b_{ij}Y_{ij}f_{Q_i}f_{D_j}$ after passing to the basis in which $Y_{ij}f_{Q_i}f_{D_j}$ is diagonalized. Contrary to the usual assumption of Yukawa anarchy, the overall size of the $b$ term depends on the misalignment of the specific anarchic Yukawa matrix relative to the set of bulk masses as flavor spurions. One expects diagrams with internal zero modes to give the dominant contributions, since these are the most sensitive to the bulk mass spectrum and hence robust against diagonalization; this intuition is confirmed by our numerical scans. 
One measure of this effect is the $1\sigma$ standard deviation from $b=0$ in a scan over random anarchic matrices \cite{Csaki:2010aj}; we use this to identify the dominant contributions to this misalignment term.

By assumption, the anarchic contribution is independent of the SM flavor sector, so there is no analogous alignment suppression to the $a$ coefficient.
However, depending on the internal modes in the loop, each diagram contributing to this term carries one of two possible independent flavor spurions that can be built out of the Yukawa matrices that may enter this product: $Y^\dag_u Y_u Y_d^\dag$ and $Y^\dag_d Y_d Y^\dag_d$. These matrices may have arbitrary relative phase, so the two terms may add either constructively or destructively. The misalignment contribution is a third independent flavor spurion, which also carries a relative phase dependent on the particular choice of parameters.


We express the anarchic ($a$) and misalignment ($b$) coefficients in terms of dimensionless integrals, which are defined in Appendix~\ref{app:integrals}. To explicitly demonstrate the calculation of diagrams in the 5D mixed position/momentum space formalism, we present a sample calculation of the anarchic contribution to $C_7$ in Appendix~\ref{sec:sample:calc}. The $C_8$ diagrams where a gluon is emitted from an internal gluon have integral results that are typically $\mathcal O(1)$ while the integrals for the other diagrams are typically $\mathcal O(10^{-1})$ in magnitude. Note that the contribution to $a$ from each diagram matches what is expected from a naive dimensional analysis. This is in contrast to the analogous calculation for $\mu\to e\gamma$, where the leading diagrams are smaller than the naive estimated size. There are thus no problems with the two-loop contribution yielding a  larger contribution than expected from the perturbative expansion.

Below we present the calculation for the right-to-left chirality (unprimed) Wilson coefficients $\Delta C_{7,8}$ for $b\to q$; the left-to-right chirality (primed) Wilson coefficients are obtained by Hermitian conjugation of the $q\to b$ amplitude.
The anarchic contribution to the left-to-right chirality coefficients are enhanced over the right-to-left coefficients by a factor of $f_{b_L}/f_{b_R}$, while the misalignment contribution is of the same order of magnitude. This behavior is explained qualitatively in Appendix~\ref{sec:misalignment} and demonstrated numerically in Section~\ref{sec:num}.

\subsection[Calculation of $\Delta C_7^{(\prime)}$]{\boldmath Calculation of $\Delta C_7^{(\prime)}$}

\begin{figure}[th]
  \centering
  \subfloat[Charged Goldstone loop]
		{\label{fig:C7:a}
		\begin{tikzpicture}[line width=1.5 pt, scale=1.2] 
			\draw[color=white] (-1.7,-1.7) rectangle (1.7,1.7);
			\draw[fermion] (-1.75,0) -- (-1,0);
			\draw[fermionbar] (1,0) -- (1.75,0);
			\draw[fermionbar] (-1,0) arc (180:135:1);
			\draw[fermion] (135:1) arc (135:45:1);
			\draw[fermion] (45:1) arc (45:0:1); 
			\draw[scalarnoarrow] (1,0) arc (0:-180:1);
			\draw[vector] (45:1) -- (45: 2);
			%
			\begin{scope}[rotate=135]
				\begin{scope}[shift={(1,0)}] 
					\clip (0,0) circle (.175cm);
					\draw[fermionnoarrow] (-1,1) -- (1,-1);
					\draw[fermionnoarrow] (1,1) -- (-1,-1);
				\end{scope}	
			\end{scope}
		 	 \node at (0,-1.4) {$H^-$};
           \node at (-1.5,-.35) {$D$};
           \node at (1.5,-.35) {$Q$};
           \node at (160:1.35) {$Q$};
           \node at (22:1.35) {$U$};
           \node at (90:1.35) {$U$};
           \node at (-.7,0) {$Y_d^\dag$};
           \node at (135:.6) {$Y_u$};
           \node at (.7,0) {$Y^{\dagger}_u$};
		\end{tikzpicture}
	}                
   \qquad
  \subfloat[Gluon ($G_\mu$ or $G_5$) loops with a single mass insertion]
		{\label{figs:C7:b}
		\begin{tikzpicture}[line width=1.5 pt, scale=1.2] 
		\draw[color=white] (-1.7,-1.7) rectangle (1.7,1.7);
		\draw[fermion] (-1.75,0) -- (-1,0);
		\draw[fermionbar] (1,0) -- (1.75,0);
		\draw[fermion] (-1,0) arc (180:135:1);
		\draw[fermionbar] (135:1) arc (135:45:1);
		\draw[fermionbar] (45:1) arc (45:0:1); 
		\draw[gluon] (1,0) arc (0:-180:1);
		\draw[vector] (45:1) -- (45:2);
		%
		\begin{scope}[rotate=135]
		\begin{scope}[shift={(1,0)}] 
			\clip (0,0) circle (.175cm);
			\draw[fermionnoarrow] (-1,1) -- (1,-1);
			\draw[fermionnoarrow] (1,1) -- (-1,-1);
		\end{scope}	
		\end{scope}
		%
		\node at (0,-1.4) {$G$};
		\node at (22.5:1.35) {$Q$};
		\node at (90:1.35) {$Q$};
		\node at (160:1.35) {$D$};
		\node at (-1.5,-.35) {$D$};
		\node at (1.5,-.35) {$Q$};
		\node at (135:.65) {$Y_d^\dag$};
	\end{tikzpicture}
	\quad
		\begin{tikzpicture}[line width=1.5 pt, scale=1.2] 
		\draw[color=white] (-1.7,-1.7) rectangle (1.7,1.7);
		\draw[fermion] (-2,0) -- (-1.5,0);
		\draw[fermionbar] (-1.5,0) -- (-1,0);
		\draw[fermionbar] (1,0) -- (1.75,0);
		\draw[fermionbar] (180:1) arc (180:45:1);
		\draw[fermionbar] (45:1) arc (45:0:1); 
		\draw[gluon] (1,0) arc (0:-180:1);
		\draw[vector] (45:1) -- (45:2);
		%
		\begin{scope}[shift={(-1.5,0)}] 
			\clip (0,0) circle (.175cm);
			\draw[fermionnoarrow] (-1,1) -- (1,-1);
			\draw[fermionnoarrow] (1,1) -- (-1,-1);
		\end{scope}	
		%
		\node at (0,-1.4) {$G$};
		\node at (22.5:1.35) {$Q$};
		\node at (112.5:1.35) {$Q$};
		\node at (-1.75,-.35) {$D$};
		\node at (1.5,-.35) {$Q$};
		\node at (-1.5,.4) {$Y_d^\dag$};
	\end{tikzpicture}
	}
  \caption{Leading contributions to the anarchic ($a$) and misalignment ($b$) terms of the $C_7$ Wilson coefficient. Arrows indicate SU(2)$_\text{L}$ representation; this is equivalent to labeling the chirality of the zero mode for SM fields.  Here $Q$, $U$ and $D$ denote the 5D chiral fermion fields containing the SM left-handed doublets and right-handed up and down singlets, respectively. $H^-$ is the charged component of the Higgs doublet that serves as the Goldstone boson of $W^-$ after electroweak symmetry breaking, and $G$ is the 5D gluon field.
  Additional diagrams related by exchanging the order of the mass insertion and photon emission are left implicit.}
  \label{fig:C7:dominant} 
\end{figure}
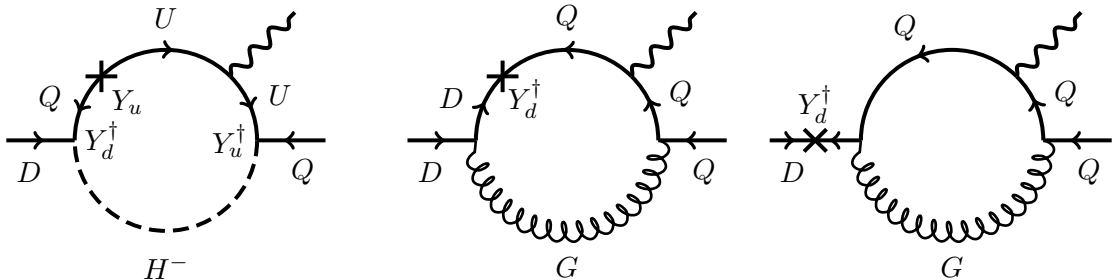

Fig.~\ref{fig:C7:dominant} shows the dominant contributions to the $C_7$ photon penguin operator. 
The RS contribution to the $b\to q\gamma$ Wilson coefficient  is
\begin{align}\label{eq:c7:expression}
\Delta C_{7} =
\frac{-vR'^2}{8\,m_b\,G_F} (V^*_{tq}V_{tb})^{-1} 
\, \sum_{ijk\ell} (U^{D_L}_{qi})^\dagger f_{Q^d_i}f_{D_{j}}\,
\left[ 	
	\displaystyle{\sum_{k,\ell}}\,a_{k\ell}Y^{u\dagger}_{ik}Y^{u}_{k\ell}Y^{d\dagger}_{\ell j}
	+ b_{ij}
Y^{d\dagger}_{ij}
			\right]U^{D_R}_{jb}.
\end{align}
$U^{D_{L,R}}$ are the rotation matrices between the 5D gauge and the light down quark mass bases.

Note that throughout our analysis we use the tree level matching condition for the 5D gauge couplings and neglect possible brane kinetic terms that may alter this matching. While this affects the misalignment contribution to $C^{(\prime)}_7$ and the calculation for $C^{(\prime)}_8$, the anarchic contribution to $C^{(\prime)}_7$, containing only one gauge coupling vertex instead of three, remains relatively unaffected. Since the latter gives the dominant contribution to the observables discussed in section \ref{sec:num}, we do not expect this assumption to have a significant impact on our predictions.

\subsubsection[$\Delta C_7$: anarchic contribution]{\boldmath $\Delta C_7$: anarchic contribution}
\label{sec:C7a}

The dominant anarchic contribution is the diagram with one mass insertion and a charged Higgs (Goldstone) in the loop, Fig.~\ref{fig:C7:a}. Note that this diagram is not present in the analogous leptonic penguin, which has a neutrino in the loop. The analogous diagram with the photon emitted off the charged Higgs propagator is found to be suppressed by a factor of $(m_WR')^2\sim10^{-2}$ due to an algebraic cancellation \cite{Csaki:2010aj}, while the one with the mass insertion on an external fermion leg is suppressed by $m_q/v$ since the external brane-to-brane fermion propagator must be a zero-mode. All other diagrams contain two additional mass insertions---necessary to obtain the required structure of a product of three Yukawas---and are therefore also suppressed by a factor of $(vR'/\sqrt{2})^2\sim10^{-2}$. Of these neglected diagrams, the next-to-leading diagrams contributing to this coefficient are gluon loops with three mass insertions, which carry a gauge coupling enhancement of $g_s^2\ln R'/R\approx 36$ but are suppressed due to the two additional mass insertions, the quark charge $(Q_d=-1/3)$, and different topologies; they are $\sim5\%$ corrections to the leading contribution. Note that these diagrams carry an independent flavor structure ($Y^\dag_d Y_d Y^\dag_d$) and can interfere either constructively or destructively with Fig.~\ref{fig:C7:a}.
%

The value for the $a$ coefficient in (\ref{eq:c7:expression}) coming from the penguin in Fig.~\ref{fig:C7:a} is a dimensionless integral whose explicit form is given in (\ref{eq:I:1MIH0}),
\begin{align}
a &= Q_u I_{C_{7a}},
\label{eq:C7:a}
\end{align}
where $Q_u=2/3$ is the charge of the internal up-type quark. 

\subsubsection[$\Delta C_7$: misalignment contribution]{\boldmath $\Delta C_7$: misalignment contribution}

The dominant misalignment contributions  come from gluon diagrams with a single mass insertion. As shown in Fig.~\ref{figs:C7:b}, this insertion can either be on an internal or external fermion line. All other diagrams contain electroweak couplings and hence are subdominant.
%
The final misalignment contribution  in (\ref{eq:c7:expression}) is
\begin{align}\label{eq:c7bcoeff}
b
= Q_d  \frac{4}{3} \left(g_s^2 \ln \frac{R'}{R}\right) I_{C_{7b}}.
\end{align}
Here $Q_d$ is the charge of the internal down-type quark, $4/3$ is a color factor, $\ln R'/R$ is a warp factor associated with bulk gauge couplings, and $I_{C_{7b}}$ is a dimensionless integral defined in (\ref{eq:I:c7b}).

\subsection[Calculation of $\Delta C_8^{(\prime)}$]{\boldmath Calculation of $\Delta C_8^{(\prime)}$}

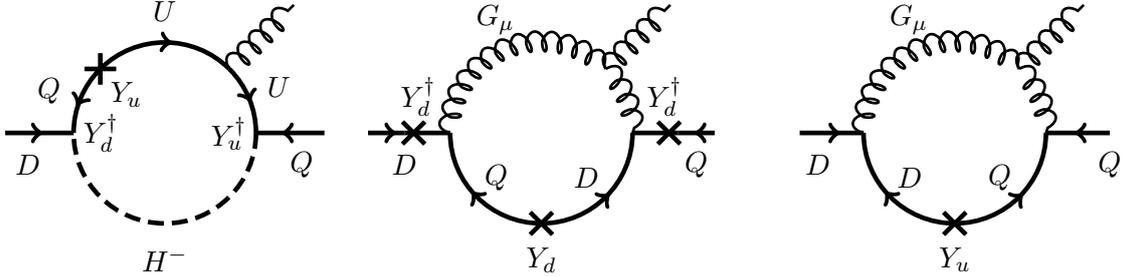
\begin{figure}[th]
  \centering
  \subfloat[Charged Goldstone and three mass insertion gluon loops. Not shown: gluon loop with two and three internal line mass insertions.]
		{\label{fig:C8:a}
		\begin{tikzpicture}[line width=1.75 pt, scale=1.2] 
			\draw[color=white] (-1.7,-1.7) rectangle (1.7,1.7);
			\draw[fermion] (-1.75,0) -- (-1,0);
			\draw[fermionbar] (1,0) -- (1.75,0);
			\draw[fermionbar] (-1,0) arc (180:135:1);
			\draw[fermion] (135:1) arc (135:45:1);
			\draw[fermion] (45:1) arc (45:0:1); 
			\draw[scalarnoarrow] (1,0) arc (0:-180:1);
			\draw[gluon] (45:1) -- (45: 2);
			%
			\begin{scope}[rotate=135]
				\begin{scope}[shift={(1,0)}] 
					\clip (0,0) circle (.175cm);
					\draw[fermionnoarrow] (-1,1) -- (1,-1);
					\draw[fermionnoarrow] (1,1) -- (-1,-1);
				\end{scope}	
			\end{scope}
		 	 \node at (0,-1.4) {$H^-$};
           \node at (-1.5,-.35) {$D$};
           \node at (1.5,-.35) {$Q$};
           \node at (160:1.35) {$Q$};
           \node at (22:1.35) {$U$};
           \node at (90:1.35) {$U$};
           \node at (-.7,0) {$Y_d^\dag$};
           \node at (135:.6) {$Y_u$};
           \node at (.7,0) {$Y^{\dagger}_u$};
		\end{tikzpicture}
		\quad
		\begin{tikzpicture}[line width=1.75 pt, scale=1.2] 
			\draw[color=white] (-1.7,-1.7) rectangle (1.7,1.7);
			\draw[fermionbar] (-1.5,0) -- (-1.9,0);
			\draw (-1.5,0) -- (-1,0);
			\draw[fermion] (1.9,0) -- (1.5,0);
			\draw (1.5,0) -- (1,0);
			\draw[gluon] (-1,0) arc (180:45:1);
			\draw[gluon] (45:1) arc (45:0:1); 
			\draw[fermionbar] (1,0) arc (0:-90:1);
			\draw[fermion] (0,-1) arc (-90:-180:1);
			\draw[gluon] (45:1) -- (45: 2);
			%
				\begin{scope}[shift={(-1.4,0)}] 
					\clip (0,0) circle (.175cm);
					\draw[fermionnoarrow] (-1,1) -- (1,-1);
					\draw[fermionnoarrow] (1,1) -- (-1,-1);
				\end{scope}	
				\begin{scope}[shift={(1.4,0)}] 
					\clip (0,0) circle (.175cm);
					\draw[fermionnoarrow] (-1,1) -- (1,-1);
					\draw[fermionnoarrow] (1,1) -- (-1,-1);
				\end{scope}	
			\begin{scope}[rotate=-90]
				\begin{scope}[shift={(1,0)}] 
					\clip (0,0) circle (.175cm);
					\draw[fermionnoarrow] (-1,1) -- (1,-1);
					\draw[fermionnoarrow] (1,1) -- (-1,-1);
				\end{scope}	
			\end{scope}
           \node at (112:1.35) {$G_\mu$};
           \node at (-135:.7) {$Q$};
           \node at (-45:.7) {$D$};
           \node at (1.35,.4) {$Y_d^\dag$};
           \node at (0,-1.4) {$Y_d$};
           \node at (-1.35,.4) {$Y_d^\dag$};
           \node at (-1.5,-.35) {$D$};
			\node at (1.7,-.35) {$Q$};
		\end{tikzpicture}
		}                
  \qquad 
    \subfloat[One mass insertion gluon loop]
		{\label{fig:C8:b}
\begin{tikzpicture}[line width=1.75 pt, scale=1.2] 
			\draw[color=white] (-1.7,-1.7) rectangle (1.7,1.7);
			\draw[fermionbar] (-1,0) -- (-1.7,0);
			\draw[fermion] (1.7,0) -- (1,0);
			\draw[gluon] (-1,0) arc (180:45:1);
			\draw[gluon] (45:1) arc (45:0:1); 
			\draw[fermionbar] (1,0) arc (0:-90:1);
			\draw[fermion] (0,-1) arc (-90:-180:1);
			\draw[gluon] (45:1) -- (45: 2);
			%
			\begin{scope}[rotate=-90]
				\begin{scope}[shift={(1,0)}] 
					\clip (0,0) circle (.175cm);
					\draw[fermionnoarrow] (-1,1) -- (1,-1);
					\draw[fermionnoarrow] (1,1) -- (-1,-1);
				\end{scope}	
			\end{scope}
           \node at (112:1.35) {$G_\mu$};
           \node at (-135:.7) {$D$};
           \node at (-45:.7) {$Q$};
           \node at (0,-1.4) {$Y_u$};
           \node at (-1.5,-.35) {$D$};
			\node at (1.7,-.35) {$Q$};
		\end{tikzpicture}
	}
  \caption{Leading contributions to the $a$ and $b$ terms of the $C_8$ Wilson coefficient following the notation of Fig.~\ref{fig:C7:dominant}. $G_\mu$ refers to only the gluon four-vector.}
  \label{fig:C8:dominant}
\end{figure}

The gluon penguin operators $C_8$ and $C_8'$ differ from their photon counterparts due to additional QCD vertices available and the magnitude of the QCD coupling, $g_{5D}^2/R=g_s^2 \ln R'/R \approx 36$. Because of this, the dominant diagrams contributing to $b\to qg$ cannot be obtained from $b\to q\gamma$ by simply replacing the photon with a gluon in the leading diagrams for $C_7^{(\prime)}$.
The general expression for $\Delta C_8$ is the same as that for $\Delta C_7$ in (\ref{eq:c7:expression}), with coefficients $a$ and $b$ coming from the diagrams shown in Fig.~\ref{fig:C8:dominant}.
%

\subsubsection[$\Delta C_8$: anarchic contribution]{\boldmath $\Delta C_8$: anarchic contribution}

There are two classes of dominant contributions to the anarchic ($a$) coefficient in $C_8^{(\prime)}$. In addition to the charged Higgs diagrams analogous to Fig.~\ref{fig:C7:a}, there are gluon diagrams with three mass insertions on the fermion lines, which are now sizable due to the size of the strong coupling constant and the three-point gauge boson vertex (as mentioned earlier, the dimensionless integral associated with this digram is $\mathcal{O}(1)$, while all other diagrams have $\mathcal{O}(0.1)$ integrals).  Of the latter class, one only needs to consider diagrams with at most one mass insertion on each external leg since sequential insertions on an external leg are suppressed by factors of $m_q R'$. 
Note that these two sets of diagrams contribute with different products of Yukawa matrices; while the Higgs diagrams are proportional to $Y_u^\dag Y_u Y_d^\dag$, the gluon diagrams are proportional to $Y_d^\dag Y_d Y_d^\dag$. Thus these two terms may add either constructively or destructively and may even add with different relative sizes if there is a hierarchy between the overall scale of the up- and down-type 5D anarchic Yukawas.
The $a$ coefficient is
\begin{align}
a = I_{C_{7a}} \oplus \frac {3}{2}\left(g_s^2 \ln \frac{R'}{R}\right)^2 \left(\frac{R'v}{\sqrt{2}}\right)^2 I_{C_{8a}}^G,
\end{align}
where we have written $\oplus$ to indicate that the two terms carry independent flavor spurions. 
Here $I_{C_{7a}}$ is the same dimensionless integral appearing in (\ref{eq:C7:a}). The second term includes color factors, warped bulk gauge couplings, and explicit mass insertions in addition to the dimensionless integral $I_{C_{8a}}$ defined in (\ref{eq:I:c8a}).

\subsubsection[$\Delta C_8$: misalignment contribution]{\boldmath $\Delta C_8$: misalignment contribution}

The single mass insertion gluon emission diagram in Fig.~\ref{fig:C8:b} gives the dominant misalignment term. Additional diagrams with the gluon emission from the quark line are suppressed by a relative color factor of $1/6$ versus $3/2$ and can be neglected. Diagrams with a scalar ($G_5$) gluon or the mass insertion on an external leg do not carry an internal fermion zero mode and become negligible after rotation to the mass basis, as discussed earlier. Diagrams with electroweak gauge bosons in the loop are suppressed due to the smaller size of the gauge coupling. The expression for the dominant diagram is
\begin{align}
b= \frac 32 \left(g_s^2 \ln \frac{R'}{R}\right) I_{C_{8b}}.
\end{align}
with $I_{C_{8b}}$ defined in (\ref{eq:I:c8b}). We have again pulled out an explicit color factor and the warped bulk gauge coupling.
%
%

\subsection{Modifications from custodial symmetry}

In models with a gauged bulk custodial symmetry, the additional matter content may also contribute to the $b\to q\gamma(g)$ transitions. By construction, boundary conditions for custodial fermions are chosen such that they have no zero modes. The misalignment (b) coefficients do not receive any significant corrections from custodial diagrams: diagrams with custodial gauge bosons are suppressed due to electroweak couplings, while those with custodial fermions do not carry internal fermion zero modes and become negligible after rotation to the mass basis. 

\begin{figure}[ht!]
  \centering
	\begin{tikzpicture}[line width=1.75 pt, scale=1.2] 
			\draw[color=white] (-1.7,-1.7) rectangle (1.7,1.7);
			\draw[fermion] (-1.75,0) -- (-1,0);
			\draw[fermionbar] (1,0) -- (1.75,0);
			\draw[fermionbar] (-1,0) arc (180:135:1);
			\draw[fermion] (135:1) arc (135:45:1);
			\draw[fermion] (45:1) arc (45:0:1); 
			\draw[scalarnoarrow] (1,0) arc (0:-180:1);
			\draw[vector] (45:1) -- (45: 2);
			%
			\begin{scope}[rotate=135]
				\begin{scope}[shift={(1,0)}] 
					\clip (0,0) circle (.175cm);
					\draw[fermionnoarrow] (-1,1) -- (1,-1);
					\draw[fermionnoarrow] (1,1) -- (-1,-1);
				\end{scope}	
			\end{scope}
		 	 \node at (0,-1.4) {$H^-$};
           \node at (-1.5,-.35) {$D_R$};
           \node at (1.5,-.35) {$Q_L^d$};
           \node at (160:1.35) {$Q_L^u$};
           \node at (22:1.40) {$U_R^{\prime(\prime)}$};
           \node at (90:1.35) {$U_R^{\prime(\prime)}$};
           \node at (-.7,0) {$Y_d^\dag$};
           \node at (135:.6) {$Y_d$};	%
           \node at (.7,0) {$Y^{\dagger}_d$}; %
		\end{tikzpicture}
		\quad
		\begin{tikzpicture}[line width=1.75 pt, scale=1.2] 
			\draw[color=white] (-1.7,-1.7) rectangle (1.7,1.7);
			\draw[fermion] (-1.75,0) -- (-1,0);
			\draw[fermionbar] (1,0) -- (1.75,0);
			\draw[fermionbar] (-1,0) arc (180:135:1);
			\draw[fermion] (135:1) arc (135:45:1);
			\draw[fermion] (45:1) arc (45:0:1); 
			\draw[scalarnoarrow] (1,0) arc (0:-180:1);
			\draw[gluon] (45:1) -- (45: 2);
			%
			\begin{scope}[rotate=135]
				\begin{scope}[shift={(1,0)}] 
					\clip (0,0) circle (.175cm);
					\draw[fermionnoarrow] (-1,1) -- (1,-1);
					\draw[fermionnoarrow] (1,1) -- (-1,-1);
				\end{scope}	
			\end{scope}
		 	 \node at (0,-1.4) {$H^-$};
           \node at (-1.5,-.35) {$D_R$};
           \node at (1.5,-.35) {$Q_L^d$};
           \node at (160:1.35) {$Q_L^u$};
           \node at (22:1.40) {$U_R^{\prime(\prime)}$};
           \node at (90:1.35) {$U_R^{\prime(\prime)}$};
           \node at (-.7,0) {$Y_d^\dag$};
           \node at (135:.6) {$Y_d$};
           \node at (.7,0) {$Y^{\dagger}_d$};
		\end{tikzpicture}
		\quad
		\begin{tikzpicture}[line width=1.75 pt, scale=1.2] 
			\draw[color=white] (-1.7,-1.7) rectangle (1.7,1.7);
			\draw[fermionbar] (-1.5,0) -- (-1.9,0);
			\draw (-1.5,0) -- (-1,0);
			\draw[fermion] (1.9,0) -- (1.5,0);
			\draw (1.5,0) -- (1,0);
			\draw[gluon] (-1,0) arc (180:45:1);
			\draw[gluon] (45:1) arc (45:0:1); 
			\draw[fermionbar] (1,0) arc (0:-90:1);
			\draw[fermion] (0,-1) arc (-90:-180:1);
			\draw[gluon] (45:1) -- (45: 2);
			%
				\begin{scope}[shift={(-1.4,0)}] 
					\clip (0,0) circle (.175cm);
					\draw[fermionnoarrow] (-1,1) -- (1,-1);
					\draw[fermionnoarrow] (1,1) -- (-1,-1);
				\end{scope}	
				\begin{scope}[shift={(1.4,0)}] 
					\clip (0,0) circle (.175cm);
					\draw[fermionnoarrow] (-1,1) -- (1,-1);
					\draw[fermionnoarrow] (1,1) -- (-1,-1);
				\end{scope}	
			\begin{scope}[rotate=-90]
				\begin{scope}[shift={(1,0)}] 
					\clip (0,0) circle (.175cm);
					\draw[fermionnoarrow] (-1,1) -- (1,-1);
					\draw[fermionnoarrow] (1,1) -- (-1,-1);
				\end{scope}	
			\end{scope}
           \node at (112:1.35) {$G_\mu$};
           \node at (-135:.7) {$Q$};
           \node at (-45:.7) {$D'$};
           \node at (1.35,.4) {$Y_d^\dag$};
           \node at (0,-1.4) {$Y_d$};
           \node at (-1.35,.4) {$Y_d^\dag$};
           \node at (-1.5,-.35) {$D$};
			\node at (1.7,-.35) {$Q$};
		\end{tikzpicture}
  \caption{Additional custodial diagrams contributing to the $C_7$ and $C_8$ coefficients.}
  \label{fig:custodial}
\end{figure}
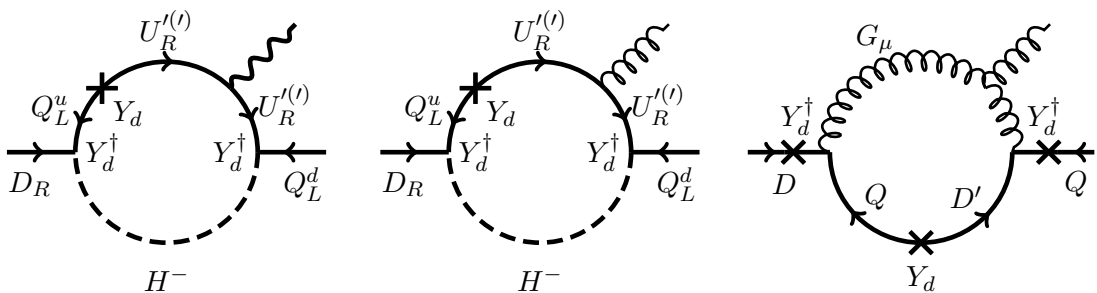

The leading custodial contributions to the anarchic (a) coefficients are shown in Fig.~\ref{fig:custodial}; these are the same diagrams that contribute to the anarchic ($a$) terms of the $C_7$ and $C_8$ Wilson coefficients and now appear with additional custodial fermions, denoted by $U'$, $U''$, and $D'$. These are the only custodial diagrams that give contributions comparable to those in Fig.~\ref{fig:C7:dominant} and Fig.~\ref{fig:C8:dominant}. The remaining diagrams consist of $W$ and $Z$ loops, which, as mentioned earlier, are suppressed by a factor of $\sim10^{-2}$ relative to the Higgs loops due to the two additional mass insertions, and remain negligible despite the larger multiplicity due to the extended electroweak sector. 

Since the custodial fermions $U'$, $U''$, and $D'$ have the same IR boundary condition as their SM counterparts but the opposite UV boundary condition, and since the localization of the Higgs pulls the loop towards the IR brane, the contribution of these custodial diagrams is well-approximated by the contributions of their SM counterparts. Since the minimal model diagrams are dominated by the KK fermion contribution, it is reasonable that the custodial modes should contribute approximately equally to the process.

Observe that each of these custodial contributions is proportional to $Y_d^\dag Y_d Y_d^\dag$. In particular, the custodial Higgs diagrams carry a flavor structure that is independent of that of their minimal model counterparts. Also, note that the $U'$ and $U''$ couplings to the charged Higgs come with a factor of $1/\sqrt{2}$ while the $D'$ coupling to the Higgs does not \cite{Blanke:2008zb}. 
Thus the additional custodial diagrams contribute an analytic structure that is nearly identical to the minimal model diagrams except for the Yukawa matrices, which now come with the product $Y_d^\dag Y_d Y_d^\dag$. 
Since this is independent of the $Y_d^\dag Y_u Y_u^\dag$ flavor spurion in the minimal model diagrams, the addition of the custodial diagrams generically enhances the penguin amplitude by less than the factor of two that one would obtain in the limit $Y_d=Y_u$.
%
%
%
This shows that while custodial symmetry can be used to suppress tree-level flavor changing effects in RS models, this comes at the cost of generically enhancing loop-level flavor processes.

\section{\boldmath Radiative $B$ decays}
\label{sec:BXsdgam}

We now examine the physical observables most directly related to the parton-level $b\to q (\gamma,g)$ operators derived above: $B$ meson decays with an on-shell photon.

\subsection[The $B\to X_{s,d}\gamma$ decay]{\boldmath The $B\to X_{s,d}\gamma$ decay}

The SM predictions for the inclusive decays $B\to X_{s,d} \gamma$ are
\cite{Misiak:2006zs,Crivellin:2011ba} 
\begin{equation}
\text{Br}(B\to X_s\gamma)_\text{SM} = (3.15\pm 0.23)\cdot 10^{-4}\,,\qquad \langle \text{Br}(B\to X_d\gamma)\rangle_\text{SM} = (15.4^{+2.6}_{-3.1})\cdot 10^{-6}\,.
\end{equation}
These can be compared to the measured values \cite{Asner:2010qj}
\begin{equation}
\text{Br}(B\to X_s\gamma)_\text{exp} = (3.55\pm 0.27)\cdot 10^{-4}\,,\qquad \langle \text{Br}(B\to X_d\gamma)\rangle_\text{exp} = (14\pm5)\cdot 10^{-6}\,.
\end{equation}
Here $\langle \text{Br}(B\to X_d\gamma)\rangle$ refers to the CP averaged branching ratio in which the hadronic uncertainties cancel to a large extent \cite{Benzke:2010js}. We have extrapolated the experimental value for $\langle \text{Br}(B\to X_d\gamma)\rangle$ to the photon energy cut $E_\gamma > 1.6\,\text{GeV}$ used for the theory prediction.

Rather than performing an extensive error analysis, we simply require the new RS contributions to fulfill the constraints
\begin{eqnarray}
\Delta \text{Br}(B\to X_s\gamma) &=&  \phantom{\langle}\text{Br}(B\to X_s\gamma)_\text{exp}\phantom{\rangle}-\phantom{\langle}\text{Br}(B\to X_s\gamma)_\text{SM}\phantom{\rangle} = (0.4 \pm 0.7)\cdot 10^{-4}\,,\label{eq:bsg-const}\\
\Delta \text{Br}(B\to X_d\gamma) &=& \langle \text{Br}(B\to X_d\gamma)\rangle_\text{exp}-\langle \text{Br}(B\to X_d\gamma)\rangle_\text{SM} = -(1 \pm 11)\cdot 10^{-6}\,.\label{eq:bdg-const}
\end{eqnarray}
Neglecting all uncertainties associated with NP contributions, these constraints represent the $2\sigma$ ranges when combining experimental and theoretical uncertainties in quadrature.
Although the data and prediction for $B\to X_d\gamma$ are currently less precise than those for $B\to X_s\gamma$, an important and partly complementary constraint can be obtained from the former decay, as recently pointed out in \cite{Crivellin:2011ba}. Since the data for $B\to X_d\gamma$ lie slightly below the SM prediction, $\Delta \text{Br}(B\to X_d\gamma)<0$ is somewhat favored, leaving little room for NP contributing to $C'_{7}$. In contrast, a positive NP contribution to $\text{Br}(B\to X_s\gamma)$ is welcome to bring the theory prediction closer to the data. We note that if the tree level values for the CKM parameters are used instead of the SM best fit values, the predicted central value for $\langle \text{Br}(B\to X_d\gamma)\rangle_\text{SM}$ rises to about $19\cdot 10^{-6}$, increasing the tension with the data.

\subsection[Master formula for $\text{Br}(B\to X_s\gamma)$]{\boldmath Master formula for $\text{Br}(B\to X_s\gamma)$}

Following the strategy of \cite{Buras:2011zb,Buras:2010pi,Blanke:2011ry}, which use the results of \cite{Misiak:2006ab}, 
the ``master formula'' for the inclusive $B\to X_s\gamma$ branching ratio in terms of the SM branching ratio, Br$_\text{SM}$, and NP contributions to the Wilson coefficients is
\begin{equation}\label{eq:Brbsg}
\text{Br}(B\to X_s\gamma) 
	= 
	\text{Br}_\text{SM} 
		+ 0.00247 \big[ 
			|\Delta C_7(\mu_b)|^2 
			+ | \Delta C'_{7}(\mu_b) |^2 
			-0.706\, \text{Re}( \Delta C_7(\mu_b) ) 
			\big]\,.
\end{equation}
%
%
%
The RS contributions to $\Delta C^{(\prime)}_{7}(\mu_b)$ are obtained from the RG evolution of $\Delta C_7^{(\prime)}$ and $\Delta C_8^{(\prime)}$, calculated in Section~\ref{sec:penguins} at the high scale $M_\text{KK} = 2.5\,\text{TeV}$, down to the $B$ scale, $\mu_b = 2.5\,\text{GeV}$,
\begin{equation}\label{eq:RGrunning}
\Delta C^{(\prime)}_{7}(\mu_b) = 
	0.429\, \Delta C^{(\prime)}_{7}(M_\text{KK}) 
	+ 0.128\, \Delta C^{(\prime)}_{8}(M_\text{KK})\,.
\end{equation}
All known SM non-perturbative contributions have been taken into account while the RS contribution is included at leading order neglecting uncertainties. This approach is an approximation to studying the effects of RS physics on the decay in question; however, in view of the other uncertainties involved---such as the the mass insertion approximation and taking into account only the leading diagrams---this approach gives sufficiently accurate results to estimate the size of RS contributions. A more accurate and detailed analysis is beyond the scope of our analysis and, in our view, premature before the discovery of RS KK modes.

\subsection[Master formula for $\langle \text{Br}(B\to X_d\gamma) \rangle$]{\boldmath Master formula for $\langle \text{Br}(B\to X_d\gamma) \rangle$}

A master formula can be obtained in a similar manner for the CP-averaged $B\to X_d\gamma$ branching ratio. Using the expressions collected in \cite{Crivellin:2011ba,Blanke:2011ry,Hurth:2003dk} we find
\begin{eqnarray}
\langle \text{Br}(B\to X_d\gamma) \rangle 
	&=&
	\langle \text{Br}_\text{SM}\rangle + 10^{-5} \Big[ 
		1.69\, \left(|\Delta C_7|^2+|\Delta C'_{7}|^2\right) 
		+ 0.24\, \left(|\Delta C_8|^2+|\Delta C'_{8}|^2\right) \nonumber\\
		&&\phantom{\text{Br}_\text{SM}\rangle + 10^{-5}} 
		+ 1.06\, \text{Re}\left[\Delta C_7\Delta C_8^*+\Delta C'_{7}\Delta C_8'^*\right] - 3.24\, \text{Re}(\Delta C_7 ) \nonumber\\
		&&\phantom{\text{Br}_\text{SM}\rangle + 10^{-5}} 
		- 0.16\,  \text{Im}(\Delta C_7)
		- 1.03\, \text{Re}(\Delta C_8)
		- 0.04\, \text{Im}(\Delta C_8)\Big],
\end{eqnarray}
where all of the RS contributions to the $b\to d$ Wilson coefficients $\Delta C^{(\prime)}_{7,8}$ are evaluated at $M_\text{KK}$.

\subsection{Analytic estimate of constraints}\label{sec:analytic:estimates:bsgam}

Assuming anarchic Yukawa couplings, one may estimate the size of the RS contributions to the Wilson coefficients in terms of the anarchic coefficients in Section~\ref{sec:C7a},
\begin{eqnarray}
|\Delta C_7(M_\text{KK})^{b\to s,d\gamma}| 
	& \sim&  \frac{1}{4\sqrt{2}G_F} a Y_*^2 R'^2 \phantom{\frac{m_b}{m_b |V_{ts}|^2} }
	\;\sim\; 0.015\, a Y_*^2 \left( \frac{R'}{1\,\text{TeV}^{-1}} \right)^2 \,,\\
|\Delta C'_{7}(M_\text{KK})^{b\to s\gamma} |
	&\sim&  \frac{1}{4\sqrt{2} G_F} a Y_*^2 R'^2 \frac{m_s}{m_b |V_{ts}|^2} 
	\;\sim\; 0.18\, a Y_*^2 \left( \frac{R'}{1\,\text{TeV}^{-1}} \right)^2\,,\label{eq:C7's-estimate}\\
|\Delta C'_{7}(M_\text{KK})^{b\to d\gamma} |
	&\sim&  \frac{1}{4\sqrt{2} G_F} a Y_*^2 R'^2 \frac{m_d}{m_b |V_{td}|^2} 
	\;\sim\; 0.20\, a Y_*^2 \left( \frac{R'}{1\,\text{TeV}^{-1}} \right)^2\,,\label{eq:C7'd-estimate}\
\end{eqnarray}
where we neglect the misalignment contributions. Here $Y_*$ is the average size of the anarchic Yukawa couplings $Y_{ij}$ which we assume to be equal for $Y_u$ and $Y_d$.

Generically the contribution to the chirality-flipped operator $C'_{7}$ is larger than the one to $C_7$ by more than an order of magnitude. This is a direct consequence of the hierarchical pattern of quark masses and CKM angles: in order to fit the observed spectrum, the left-handed $b_L$ quark has to be localized close to the IR brane, and consequently its flavor violating interactions are far more pronounced than those of the right-handed $b_R$.

Neglecting the subdominant contributions from $\Delta C_7$ and $\Delta C^{(\prime)}_{8}$, we can constrain the size of $\Delta C'_7$ by making use of the data on $\text{Br}(B\to X_s\gamma)$ and $\langle \text{Br}(B\to X_d\gamma)\rangle$. We
obtain the following constraints from the master formulas and the experimental constraints quoted above:
\begin{equation}
|\Delta C'_{7}(M_\text{KK})^{b\to s\gamma}| < 0.47\,,\qquad
|\Delta C'_{7}(M_\text{KK})^{b\to d\gamma}| < 0.77\,.
\end{equation}
Using (\ref{eq:C7's-estimate}--\ref{eq:C7'd-estimate}) and $a\sim 0.33$ we can derive an upper bound on the size of the Yukawa couplings, $Y_*$,
\begin{align}
& \frac{Y_*R'}{\text{TeV}^{-1}} < 2.8 \qquad\text{from } B\to X_s\gamma\,,\\
& \frac{Y_*R'}{\text{TeV}^{-1}} < 3.4 \qquad \text{from } B\to X_d\gamma\,,
\end{align}
For $R' = 1\,\text{TeV}^{-1}$ these are of the same order as the perturbativity bound on the Yukawa coupling \cite{Csaki:2008zd}.
We see that the generic constraint from $B \to X_s \gamma$ is slightly stronger than that from $B \to X_d \gamma$ due to the larger uncertainties in the latter case. However, since they only differ by an $\mathcal{O}(1)$ factor, in specific cases the latter constraint may be more restrictive, so one must take both processes into account when constraining the RS parameter space.

\subsection[CP asymmetry in $B \to K^*\gamma$]{\boldmath CP asymmetry in $B \to K^*\gamma$}

Like many extensions of the SM, RS generally induces large CP violating phases. 
It is thus of great interest to also study CP violation in $b\to s \gamma$ transitions. While the direct CP asymmetry in the inclusive $B\to X_s\gamma$ decay is in principle highly sensitive to NP contributions, in practice the SM contribution is dominated by long-distance physics and therefore plagued by large non-perturbative uncertainties \cite{Benzke:2010tq}. Consequently, a reliable prediction in the presence of NP is difficult.

Fortunately, a theoretically much cleaner observable is provided by the $B\to K^*\gamma$ decay. While its branching ratio is plagued by the theoretical uncertainty of the $B\to K^*$ form factors, this form factor dependence largely drops out of the time-dependent CP asymmetry \cite{Atwood:1997zr,Ball:2006cva,Ball:2006eu}
\begin{equation}
\frac{\Gamma(\bar B^0(t)\to \bar K^{*0}\gamma)- \Gamma(B^0(t)\to  K^{*0}\gamma)}{\Gamma(\bar B^0(t)\to \bar K^{*0}\gamma)+ \Gamma(B^0(t)\to  K^{*0}\gamma)} = S_{K^*\gamma}\sin(\Delta M_d t) - C_{K^*\gamma}\cos(\Delta M_d t)\,.
\end{equation}
The coefficient $S_{K^*\gamma}$ is highly sensitive to new RS contributions. At leading order it is given by \cite{Ball:2006cva,Altmannshofer:2011gn}
\begin{equation}\label{eq:SK*g}
S_{K^*\gamma} \simeq \frac{2}{|C_7|^2+|C'_{7}|^2} \text{Im} 
\left( e^{-i\phi_d}C_7C'_{7} \right)\,,
\end{equation}
where the Wilson coefficients are to be taken at the scale $\mu_b$. $\phi_d$ is the phase of $B^0$--$\bar B^0$ mixing, which has been well measured in $B^0 \to J/\psi K_S$ decays to be $\sin\phi_d = 0.67\pm 0.02$ \cite{Asner:2010qj}.

From \eqref{eq:SK*g} we see that $S_{K^*\gamma}$ is very sensitive to new phsyics in the chirality flipped operator $C'_{7}$ and vanishes in the limit $C'_{7} \to 0$. Consequently the SM prediction is suppressed by the ratio $m_s/m_b$ and is therefore very small \cite{Ball:2006eu},
\begin{equation}
S_{K^*\gamma}^\text{SM} = (-2.3\pm1.6)\%\,.
\end{equation}
Measuring a sizable CP asymmetry $S_{K^*\gamma}$ would thus not only be a clear sign of physics beyond the SM, but unambiguously indicate the presence of new right handed currents. The present experimental constraint \cite{Asner:2010qj,Ushiroda:2006fi,Aubert:2008gy},
\begin{equation}\label{eq:SK*g-exp}
S_{K^*\gamma}^\text{exp} = -16\%\pm 22\%,
\end{equation}
is still subject to large uncertainties but already puts strong constraints on NP in $b\to s$ transitions \cite{Altmannshofer:2011gn}. A significant improvement is expected soon from LHCb, and the next generation $B$ factories will reduce the uncertainty even further.

\section{\boldmath Semileptonic $B$ decays}
\label{sec:bsmumu}

Semileptonic $B$ decays such as $B\to X_s \mu^+\mu^-$ and $B\to K^*\mu^+\mu^-$ offer an interesting opportunity to not only look for deviations from the SM, but also to identify the pattern of NP contributions and therewith distinguish various NP scenarios. These decays receive contributions from semileptonic four-fermion operators $(\bar sb)(\bar\mu\mu)$ in addition to the magnetic dipole operators discussed earlier. While the dipole operators receive RS contributions first at the one-loop level as required by gauge invariance, the four fermion operators are already affected at tree level by the exchange of the $Z$ boson and the heavy electroweak KK gauge bosons.

In this section we discuss the effective Hamiltonian for $b\to s\mu^+\mu^-$ transitions. Subsequently we will review a number of benchmark observables that are relevant for the study of RS contributions.

\subsection[Effective Hamiltonian for $b\to s\mu^+\mu^-$ transitions]{\boldmath Effective Hamiltonian for $b\to s\mu^+\mu^-$ transitions}

The effective Hamiltonian for $b\to s\mu^+\mu^-$  reads
\begin{eqnarray}
\mathcal{H}_\text{eff} &=& \mathcal{H}_\text{eff}(b\to s \gamma)- \frac{G_F}{\sqrt{2}}V_{ts}^*V_{tb}
\Big[C_{9V}(\mu)Q_{9V}(\mu)+ C'_{9V}(\mu)Q'_{9V}(\mu)\nonumber \\
& &\hspace{4cm} +C_{10A}(\mu)Q_{10A}(\mu)+C'_{10A}(\mu)Q'_{10A}(\mu)\Big]+\text{h.c.}\,,\qquad
\end{eqnarray}
where we neglect the terms proportional to $V_{us}^* V_{ub}$, and
\begin{align}
&Q_{9V} = 2(\bar s\gamma_\mu P_L b)(\bar\mu\gamma^\mu\mu)&& Q'_{9V} = 2(\bar s\gamma_\mu P_R b)(\bar\mu\gamma^\mu\mu)\\
&Q_{10A} = 2(\bar s\gamma_\mu P_L b)(\bar\mu\gamma^\mu\gamma_5\mu) && Q'_{10A} = 2(\bar s\gamma_\mu P_R b)(\bar\mu\gamma^\mu\gamma_5\mu).
\end{align}
In the SM only the unprimed Wilson coefficients are relevant. At the scale $M_W$ they are given by
\begin{equation}
C_{9V}^{\text{SM}}(M_W) = \frac{\alpha}{2\pi}\left[\frac{Y_0(x_t)}{\sin^2\theta_W}-4Z_0(x_t)\right]\,\qquad
C_{10A}^{\text{SM}}(M_W) = -\frac{\alpha}{2\pi}\frac{Y_0(x_t)}{\sin^2\theta_W}
\label{eq:C9:C10:SM}
\end{equation}
where $x_t= m_t^2/M_W^2$ and the dimensionless loop functions
 $Y_0(x_t) \approx 0.94$ and $Z_0(x_t)\approx 0.65$  are explicitly written in (3.27) and (3.28) of \cite{Buras:1998raa}.

While $C^{(\prime)}_{7}$ and $C^{(\prime)}_{8}$ receive the loop-level RS contributions calculated in Section~\ref{sec:penguins}, $C^{(\prime)}_{9V}$ and  $C^{(\prime)}_{10A}$ are corrected at tree level from the new flavor-changing couplings to the $Z$ boson and the exchange of neutral electroweak gauge boson KK modes. In this analysis we only keep the leading contribution to each of these operators, i.e.\ we consider $\Delta C^{(\prime)}_{7\gamma,8G}$ at one loop and $\Delta C^{(\prime)}_{9V,10A}$ at tree level. Strictly speaking, such an approach leads to an inconsistent perturbative expansion, but it is reasonable to expect that the one loop corrections to the latter Wilson coefficients are sub-dominant with respect to the tree level contributions, and by considering only the RS tree level contribution one should still capture the dominant NP effects.

Explicit expressions for $\Delta C^{(\prime)}_{9V}$ and $\Delta C^{(\prime)}_{10A}$ can be straightforwardly obtained from \cite{Blanke:2008yr}. These expressions can be written in terms of RG invariants $\Delta Y^{(\prime)}$ and $\Delta Z^{(\prime)}$ and the coupling $\alpha$, which itself is only very weakly scale dependent above $M_W$. Thus one may use these expressions to directly write the RS contributions at the scale $M_W$,
\begin{eqnarray}
\Delta C_{9V} &=& \frac{\alpha}{2\pi}\left[\frac{\Delta Y_s}{\sin^2\theta_W}-4\Delta Z_s\right]\\
\Delta C'_{9V} &=& \frac{\alpha}{2\pi}\left[\frac{\Delta Y'_s}{\sin^2\theta_W}-4\Delta Z'_s\right]\\
\Delta C_{10A} &=& -\frac{\alpha}{2\pi}\frac{\Delta Y_s}{\sin^2\theta_W}\\
\Delta C'_{10A} &=& -\frac{\alpha}{2\pi}\frac{\Delta Y'_s}{\sin^2\theta_W}
\end{eqnarray}
The functions $\Delta Y^{(\prime)}$ and $\Delta Z^{(\prime)}$ are given by
\begin{eqnarray}
\Delta Y_s &=& -\frac{1}{V_{ts}^* V_{tb}}\sum_{X} \frac{\Delta_L^{\mu\mu}(X)-\Delta_R^{\mu\mu}(X)}{4 M_X^2 g_\text{SM}^2} \Delta_L^{bs}(X)\,,\\
\Delta Y'_s &=& -\frac{1}{V_{ts}^* V_{tb}}\sum_{X} \frac{\Delta_L^{\mu\mu}(X)-\Delta_R^{\mu\mu}(X)}{4 M_X^2 g_\text{SM}^2} \Delta_R^{bs}(X)\,,\\
\Delta Z_s &=& \phantom{+} \frac{1}{V_{ts}^* V_{tb}}\sum_{X} \frac{\Delta_R^{\mu\mu}(X)}{8 M_X^2 g_\text{SM}^2\sin^2\theta_W}\Delta_L^{bs}(X)\,,\\
\Delta Z'_s &=& \phantom{+} \frac{1}{V_{ts}^* V_{tb}}\sum_{X} \frac{\Delta_R^{\mu\mu}(X)}{8 M_X^2 g_\text{SM}^2\sin^2\theta_W}\Delta_R^{bs}(X)\,.
\end{eqnarray}
Here the summation runs over $X = Z, Z^{(1)},A^{(1)}$ in the minimal model and over $X = Z, Z_H, Z', A^{(1)}$ in the custodial model. The flavor violating 4D fermion gauge boson couplings $\Delta^{ij}_{L,R}(X)$ depend on the overlap of the fermion profile with the corresponding gauge boson profile. Their explicit form depends on both the fermion and gauge boson mixing matrices. The explicit expressions are complicated and unilluminating, hence we do not quote them here but refer the reader to appendix A of \cite{Blanke:2008yr}. Furthermore 
\begin{equation}
g_\text{SM}^2 = \frac{G_F}{\sqrt{2}} \frac{\alpha}{2\pi \sin^2\theta_W}\,.
\end{equation}

The tree level contributions to $b \to s \mu^+\mu^-$ transitions in the minimal RS model are evaluated in \cite{Bauer:2009cf} without making the approximations of taking into account only the first KK modes or treating the Higgs vacuum expectation value as a perturbation. 
In this paper we are mainly interested in the effects of $\sim 2.5\,\text{TeV}$ KK modes. As these are ruled out in the minimal model by precision electroweak constraints, we focus on the phenomenological effects of the custodial RS model on these transitions.

For the study of observables related to $b\to s\mu^+\mu^-$, it is useful to introduce the effective Wilson coefficients at the scale $\mu_b$ that include the effects of operator mixing,
\begin{align}
C_7^\text{eff} =&\  (C_7^\text{eff})_\text{SM} + \Delta C_7(\mu_b)\,,\qquad & C_7^{\prime\text{eff}} =&\  (C_7^{\prime\text{eff}})_\text{SM} +\Delta C'_7(\mu_b)
\label{eq:C7:eff}\,,\\
C_{9V}^\text{eff}(q^2) =&\  (C_{9V}^\text{eff})_\text{SM}(q^2) + \frac{2\pi}{\alpha}\Delta C_{9V}\,,\qquad & C_{9V}^{\prime\text{eff}} =&\  \frac{2\pi}{\alpha}\Delta C'_{9V}
\label{eq:C9:eff}\,,\\
C_{10A}^\text{eff} =&\ (C_{10A}^\text{eff})_\text{SM} + \frac{2\pi}{\alpha}\Delta C_{10A}\,,\qquad & C_{10A}^{\prime\text{eff}} =&\  \frac{2\pi}{\alpha}\Delta C'_{10A}
\label{eq:C10:eff}\,.
\end{align}
The SM values of the effective Wilson coefficients can be found in Table~2 of \cite{Altmannshofer:2008dz}, which also gives the $q^2$ dependence of $(C_{9V}^\text{eff})_\text{SM}(q^2)$ in terms of a linear combination of the other Wilson coefficients. While in principle all contributions have to be taken at the scale $\mu_b$, the NP contributions to $C_{9V,10A}^{(\prime)}$ are invariant under renormalization group evolution. 

With these effective Wilson coefficients at the $B$ scale, we are now equipped to study observables in $b\to s\mu^+\mu^-$ transitions. While this system offers a plethora of observables for study, a detailed analysis of all of them is beyond the scope of this paper, and we concentrate on studying a few benchmark observables that are particularly relevant for RS physics. A numerical analysis is presented in Section~\ref{sec:num}. 

In passing we would like to remark on the pattern of contributions to $C^{(\prime)}_{9V,10A}$ in the custodial model, as pointed out in \cite{Blanke:2008yr}. Due to the suppression of flavor violating $Z d_L^i \bar d_L^j$ couplings by the discrete $P_{LR}$ symmetry, the main contributions arise in the primed Wilson coefficients $C'_{9V,10A}$, which are absent in the SM. Since the right-handed $b$ quark, localized significantly further away from the IR brane than the left-handed one, is far less sensitive to flavor violating effects introduced by the RS KK modes, the RS effects in $Y_s^{(\prime)},Z_s^{(\prime)}$ turn out to be rather small (typically below 10\%). This pattern is very different from the minimal model, where the $P_{LR}$ suppression mechanism is absent and large tree level flavor violating $Z$ couplings to left-handed down-type quarks are present.

\subsection{Benchmark observables}

\subsubsection[$\text{Br}(B\to X_s\mu^+\mu^-)$]{\boldmath $\text{Br}(B\to X_s\mu^+\mu^-)$}

For very low lepton invariant mass $q^2 \to 0$, the $B\to X_s\mu^+\mu^-$ transition is completely dominated by the photon pole and doesn't provide any new insight with respect to the $B\to X_s\gamma$ decay discussed above. Furthermore, in the intermediate region $6\,\text{GeV}^2 < q^2 < 14.4\,\text{GeV}^2$ the sensitivity to NP is very small, as the decay rate in this region is completely dominated by charm resonances. Hence one usually restricts oneself to either the {low $q^2$ region} $1\,\text{GeV}^2 < q^2 < 6\,\text{GeV}^2$, or the {high $q^2$ region} $ q^2 > 14.4\,\text{GeV}^2$. In what follows we will consider only the low $q^2$ region. While the high $q^2$ region is potentially interesting since it exhibits a small tension between SM prediction \cite{Hurth:2008jc} and experimental data \cite{Aubert:2004it,Iwasaki:2005sy}, it is far less sensitive to NP in 
$C^{(\prime)}_{7}$, which is the main focus of this study. In the custodial RS model, the tension in the high $q^2$ region cannot be resolved since the new contributions to $C_{9V,10A}^{(\prime)}$ are generally small \cite{Blanke:2008yr}. In addition, the high $q^2$ region is subject to larger theoretical uncertainties.

In the low $q^2$ region, adapting the formulae of \cite{DescotesGenon:2011yn} to the more general case of complex NP contributions, we find
\begin{equation}
\text{Br}(B\to X_s\mu^+\mu^-)^{\text{low }q^2 } = \text{Br}(B\to X_s\mu^+\mu^-)_\text{SM}^{\text{low } q^2} + \Delta \text{Br}(B\to X_s\mu^+\mu^-)^{\text{low }q^2 }
\end{equation}
with the NNLL prediction \cite{Huber:2005ig} 
\begin{equation}
\text{Br}(B\to X_s\mu^+\mu^-)_\text{SM}^{ \text{low }q^2 } = (15.9\pm 1.1)\cdot 10^{-7}
\end{equation}
and the NP contribution 
\cite{DescotesGenon:2011yn}
\begin{eqnarray}
\Delta \text{Br}(B\to X_s\mu^+\mu^-)^{\text{low } q^2} &\simeq& 10^{-7}\cdot \Big[
-0.517\,\text{Re}(\Delta C_7(\mu_b)) -0.680\, \text{Re}(\Delta C'_{7}(\mu_b))\qquad\qquad\nonumber\\
&&{}\qquad\quad +2.663\, \text{Re}(\delta C_{9V}) -4.679 \,\text{Re}(\delta C_{10A}) \nonumber\\
&&{}\qquad\quad + 27.776 \left(|\Delta C_7(\mu_b)|^2+ |\Delta C'_{7}(\mu_b)|^2\right) \nonumber\\
&&{}\qquad\quad + 0.534 \left(|\delta C_{9V}|^2+ |\delta C'_{9V}|^2\right)\nonumber\\
&&{}\qquad\quad + 0.543 \left(|\delta C_{10A}|^2+ |\delta C'_{10A}|^2\right)\nonumber\\
&&{}\qquad\quad + 4.920 \,\text{Re}\left(\Delta C_7(\mu_b) \delta C_{9V}^* + \Delta C'_{7}(\mu_b) \delta C_{9V}^{\prime*}\right)\Big]\,,
\end{eqnarray}
where we defined
\begin{equation}
\delta C_i = \frac{2\pi}{\alpha}\Delta C_i.
\end{equation}
Note that we dropped all interference terms between unprimed and primed contributions since they are suppressed by a factor $m_s/m_b$ and therefore small. The only exception is the term linear in $\Delta C'_{7}$, which receives a large numerical enhancement factor, and is therefore non-negligible; hence we keep it in our analysis.

The measurements of BaBar \cite{Aubert:2004it} and Belle \cite{Iwasaki:2005sy} yield the averaged value
\begin{equation}
\text{Br}(B\to X_s\mu^+\mu^-)^{\text{low }q^2 }_\text{exp} = (16.3 \pm 5.0)\cdot 10^{-7}.
\end{equation}
As  LHCb is not well suited for performing inclusive measurements, a significant reduction of uncertainties will only be feasible at the next generation $B$ factories Belle-II and SuperB \cite{Aushev:2010bq,Bona:2007qt,OLeary:2010af,Meadows:2011bk}.

\subsubsection[$B\to K^{0*}(\to \pi K)\mu^+\mu^-$]{\boldmath $B\to K^{0*}(\to \pi K)\mu^+\mu^-$}

While the inclusive $B \to X_s \mu^+\mu^-$ mode is theoretically very clean, such measurements are experimentally challenging, and competitive results (in particular for angular distributions) will not be available before the Belle II and SuperB era \cite{
	Aushev:2010bq,
	Bona:2007qt,
	OLeary:2010af,
	Meadows:2011bk}. For this reason, exclusive decay modes have received well-deserved attention. An especially interesting decay is $B \to K^*(\to K\pi)\mu^+\mu^-$, where a plethora of angular observables can be studied thanks to the four-body final state \cite{
Altmannshofer:2011gn,	
Altmannshofer:2008dz,	
Kruger:1999xa, 			
Egede:2008uy, 			
Egede:2010zc,				
Matias:2012xw,
Bobeth:2010wg,
Bobeth:2011gi}. 
These can provide detailed information on the operator and flavor structure of the underlying NP scenario.

The downside is that many $B \to K^*(\to K\pi)\mu^+\mu^-$ observables, such as the branching ratio and differential decay distribution, are plagued by large theoretical uncertainties in the determination of the $B \to K^*$ matrix elements governed by long-distance non-perturbative QCD dynamics. These matrix elements are most conveniently described by a set of seven form factors. Presently, the best predictions for these form factors at large final state meson $K^*$ energies, i.e.\ small lepton invariant mass $q^2$, stem from QCD sum rules at the light cone \cite{Colangelo:2000dp}. Furthermore, non-factorizable corrections are calculated using QCD factorization, which is only valid in the low $q^2$ regime.\footnote{Significant progress has recently been made on the form factor predictions in the large $q^2$ region \cite{Beylich:2011aq, Beneke:2001at, Grinstein:2004vb, Beneke:2004dp}; nevertheless we will not consider this kinematic regime since it is less sensitive to NP entering $C^{(\prime)}_7$ than the low $q^2$ region.}
On the other hand, as mentioned above, at very low $q^2< 1\,\text{GeV}^2$ the $b\to s\mu^+\mu^-$ transition is dominated by the $C_7^{(\prime)}$ contributions due to the infrared photon pole and therefore does not provide any insight beyond what is already obtained from $b\to s\gamma$. Consequently, we henceforth restrict our attention to the range $1\,\text{GeV}^2 \le q^2 \le 6\,\text{GeV}^2$.

Fortunately, it is possible to partly circumvent the theoretical uncertainties by studying angular observables that are less dependent on the form factors in question. Detailed analyses of their NP sensitivity and discovery potential have been performed by various groups, both model-independently and within specific NP scenarios \cite{
Altmannshofer:2011gn,	
Altmannshofer:2008dz,	
Kruger:1999xa, 			
Egede:2008uy, 			
Egede:2010zc,				
Matias:2012xw}. We leave such a detailed analysis in the context of RS models for future work. We focus instead on two benchmark observables, the forward backward asymmetry $A_\text{FB}$, which is experimentally well constrained, and the transverse asymmetry $A_T^{(2)}$, which offers unique sensitivity to NP in the primed Wilson coefficients.

We note that the recently measured CP asymmetry $A_9$ \cite{Aaltonen:2011ja, LHCb-CONF-2012-008}, as defined in \cite{Altmannshofer:2008dz, Bobeth:2008ij}, is also very sensitive to NP in $C'_7$ and therefore is in principle an interesting observable to look for RS effects. Because it is sensitive to the phase of $C'_7$, it yields partly complementary information with respect to the CP conserving transverse asymmetry $A_T^{(2)}$. 
%
%
Although this CP asymmetry is theoretically very clean, contrary to those studied in  \cite{Egede:2010zc}, we leave a detailed study within RS for future work.

\paragraph{Forward backward asymmetry}

The forward-backward asymmetry $A_\text{FB}$ in $B\to K^*\mu^+\mu^-$ decays is defined by
\begin{equation}
A_\text{FB}(q^2) = \frac{1}{d\Gamma/dq^2}\left(\int_0^1 d(\cos\theta_\mu)\,\frac{d^2\Gamma}{dq^2d(\cos\theta_\mu)} - \int_{-1}^0 d(\cos\theta_\mu)\,\frac{d^2\Gamma}{dq^2d(\cos\theta_\mu)} \right)\,,
\end{equation}
where $\theta_\mu$ is the angle between the $K^*$ momentum and the relative momentum of $\mu^+$ and $\mu^-$. 
$A_\text{FB}$ has recently received a lot of attention as data from BaBar, Belle, and the Tevatron seem to indicate a deviation from the SM, albeit with low statistical significance \cite{:2008ju,Ishikawa:2006fh,Aaltonen:2011ja}. On the other hand, recent LHCb data \cite{LHCb-CONF-2011-038} show excellent agreement with the SM prediction, and as uncertainties are presently dominated by statistics, an improved measurement should be available soon.  

A precise theoretical determination of $A_\text{FB}$ is appealing since it offers a sensitive probe of the helicity of NP contributions. 
To leading order, the forward backward asymmetry is proportional to \cite{Bobeth:2008ij}
\begin{equation}\label{eq:AFB}
A_\text{FB}(q^2) \propto \text{Re}\left[\left(C_{9V}(q^2)+\frac{2m_b^2}{q^2}C_7\right)C_{10A}^* - \left(C_{9V}^\prime+\frac{2m_b^2}{q^2}C_7^\prime\right)C_{10A}^{\prime *}\right]
\,,
\end{equation}
where we dropped the superscript ``eff'' for the effective Wilson coefficients at the scale $\mu_b$, (\ref{eq:C7:eff}--\ref{eq:C10:eff}). From \eqref{eq:AFB} we can see explicitly that $A_\text{FB}$ does not receive contributions from the interference of different chirality operators (unprimed and primed). Consequently, with the SM contribution being the dominant effect, potential non-standard effects in $A_\text{FB}$ arise mainly from NP in $C_7$ and $C_{9V}$. On the other hand, $A_\text{FB}$ is rather insensitive to NP in the primed Wilson coefficients $C'_{7\gamma,9V,10A}$. 

$A_\text{FB}$ has been studied in the context of the minimal RS model considering only tree level contributions and omitting loop level dipole contributions to $C^{(\prime)}_{7}$ \cite{Bauer:2009cf}, where small positive contributions to $A_\text{FB}$ were found.  While $A_\text{FB}$ is very sensitive to NP effects in $C_7$, the RS dipole contributions we calculated predict rather small contributions to this Wilson coefficient. On the other hand, $A_\text{FB}$ is insensitive to $C'_{7}$, where RS effects are expected to be more pronounced over the SM. Thus the overall prediction of small deviations of $A_\text{FB}$ from the SM obtained in \cite{Bauer:2009cf} remains consistent with our calculations.
Note that the restriction to tree level RS effects is not necessarily a good approximation for observables sensitive to $C'_{7}$, such as $F_L$, which was also studied in \cite{Bauer:2009cf}. A detailed study including one-loop contributions to the dipole operators would therefore be desirable but lies beyond the scope of the present analysis.

In the custodial RS model, due to the protection of the $Z d_L^i \bar d_L^j$ vertex \cite{Blanke:2008zb},  the RS contributions to $C_{9V,10A}$ are highly suppressed, and only the new contributions to the primed operators are relevant. As $A_\text{FB}$ is insensitive to the latter Wilson coefficients, it remains very close to the SM prediction. 

We conclude that RS effects in the forward backward asymmetry $A_\text{FB}$ are generally small, so the recent data from LHCb do not pose any stringent constraint on the minimal or custodial model, the latter being even more insensitive to RS contributions. 

\paragraph{Transverse asymmetry \boldmath $A_T^{(2)}$}

The asymmetries $A_T^{(i)}$, which are introduced in 
\cite{Egede:2008uy,Kruger:2005ep}, offer a particularly good probe of NP in $b\to s\mu^+\mu^-$ transitions since at leading order they are free of any hadronic uncertainties and are given in terms of calculable short distance physics. In this paper we will restrict ourselves to the study of the asymmetry
\begin{equation}
A_T^{(2)} = \frac{|A_\perp|^2-|A_\parallel|^2}{|A_\perp|^2+|A_\parallel|^2}\,.
\end{equation}
Here $A_\perp$ and $A_\parallel$ are the transversity amplitudes \cite{Kruger:2005ep}
describing the polarization of the $K^*$ and the $\mu^+\mu^-$ pair; both are transverse with linear polarization vectors perpendicular ($\perp$) or parallel ($\parallel$) to each other. In the limit of heavy quark ($m_B \to\infty$) mass and large $K^*$ energy (small $q^2$), this asymmetry takes a particularly simple form \cite{Egede:2010zc}
\begin{equation}\label{eq:AT2}
A_T^{(2)}(q^2) = \frac{2\left[\text{Re}(C'_{10A}C_{10A}^*)+F^2\, \text{Re}(C'_{7}C_7^*) + F\, \text{Re}(C'_{7} C_{9V}^*) \right]}{|C_{10A}|^2+|C'_{10A}|^2+F^2\,(|C_7|^2+|C'_{7}|^2)+ |C_{9V}|^2+2F\,\text{Re}(C_7C_{9V}^*)}
\end{equation} 
with $F=2m_bm_B/q^2$, and we have again dropped the superscript ``eff'' from the Wilson coefficients.
In this limit it is clear that $A_T^{(2)}$ is independent of form factors and is governed only by calculable short distance physics, making this observable theoretically clean. Second, we notice that since the primed Wilson coefficients are highly suppressed in the SM, $(A_T^{(2)})_\text{SM}$ is very small. $A_T^{(2)}$ therefore offers unique sensitivity to NP entering dominantly in the primed operators $C'_{7\gamma,9V,10A}$. This asymmetry is thus a benchmark observable for discovering RS physics in $B \to K^* \mu^+\mu^-$ decays. We investigate the possible size of RS contributions to this channel in our numerical analysis in the next section.

A first measurement of $A_T^{({2})}$ by CDF \cite{Aaltonen:2011ja} is still plagued by large uncertainties. 
LHCb has recently put more stringent constraints on this asymmetry, and more precise measurements will be possible in the near future \cite{LHCb-CONF-2012-008}. With $10\,\text{fb}^{-1}$ of data, LHCb is expected to reach a sensitivity of about $\pm 0.16$.

\section{Numerical analysis}\label{sec:num}

\subsection{Strategy}

In this section we present a numerical analysis of the observables introduced in the previous sections. To this end we follow the following strategy:
\begin{enumerate}
\item
The first goal is to understand the generic pattern of effects induced by RS penguins on flavor observables. We generate a set of parameter points that satisfy the known experimental constraints from quark masses and CKM parameters. However, we do not yet impose any additional flavor bounds so as not to be biased by their impact. With these points we evaluate the new RS contributions to the Wilson coefficients $\Delta C^{(\prime)}_{7}$ and $\Delta C^{(\prime)}_{8}$ at the KK scale for both the minimal and the custodial model. Subsequently we calculate the new contributions to the branching ratios of $B\to X_{s,d}\gamma$ and analyze the constraints.

Note that the same set of parameter points is used for the minimal and the custodial model in this case, in order to minimize the sampling bias on the results obtained. 
\item
The second goal is to understand the effect of the RS penguins on the existing parameter space for realistic RS models. We restrict our attention to the custodial model, which can be made consistent with electroweak precision tests for KK scales as low as $M_\text{KK}\simeq 2.5\,\text{TeV}$. In addition to quark masses and CKM parameters, we now also impose constraints from $\Delta F = 2$ observables which are analyzed at length in \cite{Blanke:2008zb}. After evaluating the size of the effects in the $B\to X_{s,d}\gamma$ branching ratios and their constraint on the model, we study the benchmark observables outlined above, namely the CP asymmetry in $B\to K^*\gamma$, the branching ratio $\text{Br}(B\to X_s\mu^+\mu^-)$, and the transverse asymmetry $A_T^{(2)}$ in $B\to K^*\mu^+\mu^-$ decays.
\end{enumerate}
Throughout our analysis we restrict ourselves to $1/R' = 1\,\text{TeV}$, so that the lowest KK gauge bosons have a mass of $M_\text{KK}\simeq2.5\,\text{TeV}$. We note that in the minimal model such low KK masses are already excluded due to unacceptably large corrections to electroweak precision observables. However, we use the same mass scale for both the minimal and custodial models to enable a straightforward comparison of the two sets of results. Furthermore, we restrict the fundamental Yukawa couplings to lie in their perturbative regime, i.\,e.\ $|Y_{ij}|\le 3$. More details on the parameter scan can be found in \cite{Blanke:2008zb}.

\subsection{General pattern of RS contributions}

This part of the numerical analysis is dedicated to determining the size of NP effects generated by the RS KK modes in the dipole operators $C_7$, $C'_{7}$ and $C_8$, $C'_8$ mediating the $b\to (s,d)\gamma$ and $b\to (s,d)g$ transitions respectively. We advise caution when interpreting the density of points since these distributions are influenced by the details of the parameter scan performed. The qualitative features in our plots should however remain unaffected by the scanning procedures.

\begin{figure}[t!]
\begin{minipage}{7.4cm}
\includegraphics[width=\textwidth]{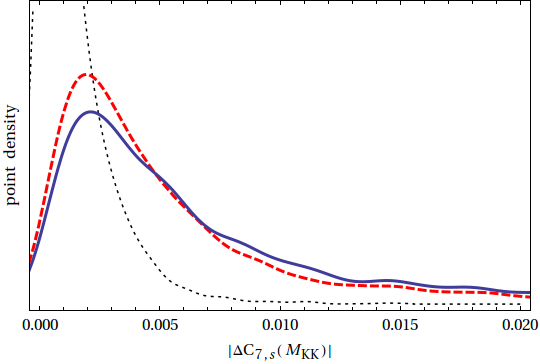}
\end{minipage}
\hfill
\begin{minipage}{7.4cm}
\includegraphics[width=\textwidth]{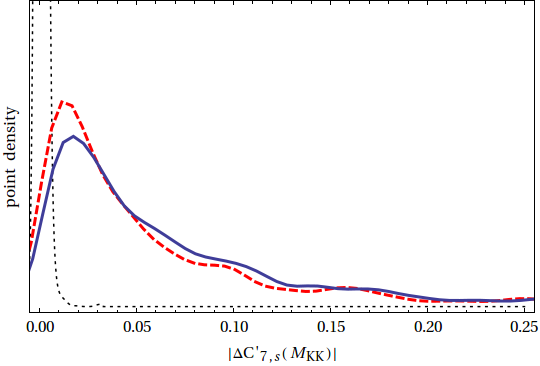}
\end{minipage}\vspace{2mm}

\begin{minipage}{7.4cm}
\includegraphics[width=\textwidth]{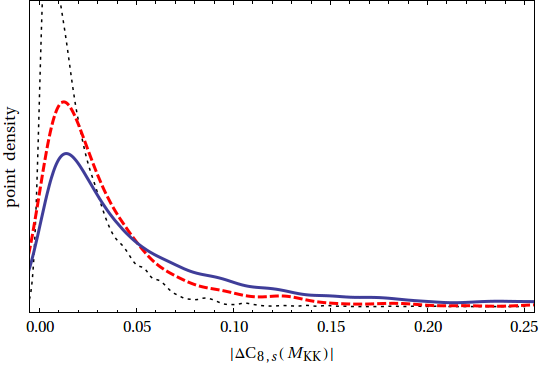}
\end{minipage}
\hfill
\begin{minipage}{7.4cm}
\includegraphics[width=\textwidth]{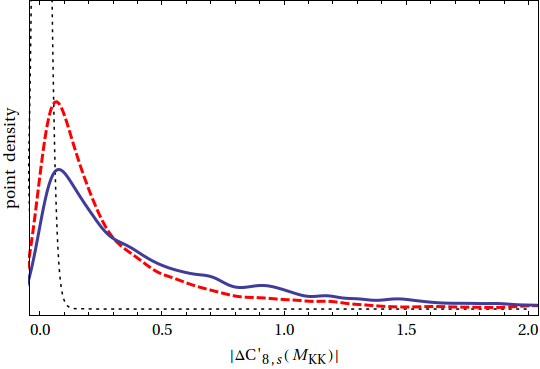}
\end{minipage}
\caption{RS contributions to the $b\to s$ Wilson coefficients $C_7(M_\text{KK})$ (upper left), $C'_{7}(M_\text{KK})$ (upper right), $C_8(M_\text{KK})$ (lower left) and  $C'_{8}(M_\text{KK})$ (lower right)
in the minimal (red, dashed) and custodial (blue, solid) models, and from the misalignment contribution alone (black, dotted).\label{fig:Cs}}
\end{figure}

The first row of Fig.~\ref{fig:Cs} shows the RS contributions to $C_7(M_\text{KK})$ and $C'_{7}(M_\text{KK})$ in the\linebreak $b\to s$ system. Observe that the total RS contribution (dashed red and solid blue distributions, corresponding to the minimal and custodial model) to the primed Wilson coefficient is typically an order of magnitude larger than the corresponding effect in the unprimed Wilson coefficient. This matches the naive expectation that the $b_L\to s_R$ transition should be enhanced relative to $b_R\to s_L$ due to the hierarchy $f_{Q_3}\gg f_{b_R}$ of fermion localizations. Furthermore the custodial contribution is somewhat enhanced relative to the minimal one, due to the additional fermion modes running in the loop.
Also shown, in black (dotted), is the contribution to $C_7(M_\text{KK})$ and $C'_{7}(M_\text{KK})$ generated by only the misalignment term, which is equal for the minimal and the custodial models. Unlike the anarchic term, this contribution is generically comparable in both cases.
This naively unexpected behavior is explained in Appendix~\ref{sec:misalignment}.
While it is subdominant but non-negligible in the case of $C_7(M_\text{KK})$, it turns out to be generally irrelevant in the case of $C'_{7}(M_\text{KK})$.

The second row of Fig.\ \ref{fig:Cs} shows the results for the gluonic penguin Wilson coefficients $C_8$ and $C'_8$. The values at the KK scale are larger than the corresponding values of $C_7$ and $C'_7$ by about an order of magnitude due to the large contribution from the diagram containing the non-Abelian $SU(3)_c$ vertex, which is absent in the $b\to s\gamma$ penguin. 
Other than that, the pattern of effects is qualitatively similar to that for $C^{(\prime)}_7$: the primed Wilson coefficient is larger than the unprimed coefficient by about an order of magnitude, and the custodial model yields somewhat bigger effects than the minimal model. Furthermore, the misalignment contributions to the unprimed and primed Wilson coefficients are again roughly comparable; consequently, its effect is negligible in $C'_8$ but can be sizable in $C_8$.

\begin{figure}
\begin{minipage}{7.4cm}
\includegraphics[width=\textwidth]{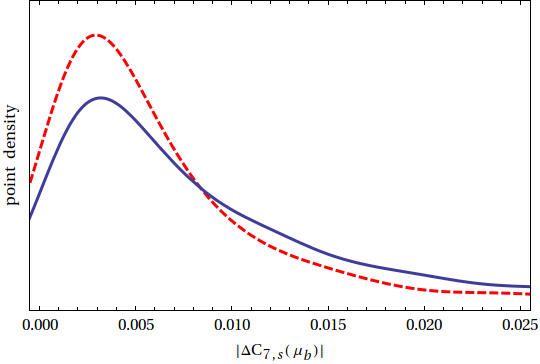}
\end{minipage}
\hfill
\begin{minipage}{7.4cm}
\includegraphics[width=\textwidth]{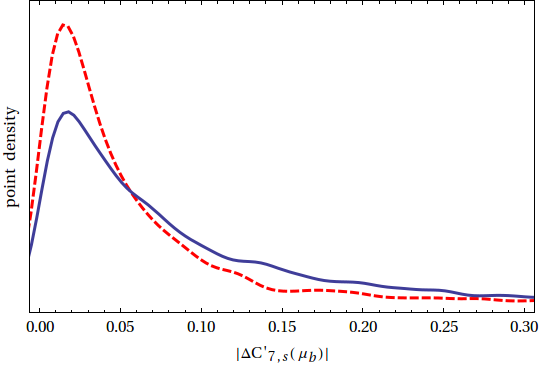}
\end{minipage}
\caption{RS contributions to the $b\to s\gamma$ 
Wilson coefficients $C_7$ (left) and $C'_{7}$ (right), evaluated at the scale $\mu_b = 2.5\,\text{GeV}$. The minimal model distribution is shown in red (dashed), and the custodial one in blue (solid). \label{fig:C7s_mb}}
\end{figure}

To facilitate comparison with other models of NP,
Fig.\ \ref{fig:C7s_mb} shows the RS contributions to the $b\to s\gamma$ Wilson coefficients $C_7$ (left) and $C'_{7}$ (right) evaluated at the scale $\mu_b = 2.5\,\text{GeV}$, i.e.\ taking into account the RG evolution and operator mixing with $C_8^{(\prime)}$. The RS contribution to $C_7$ turns out to be small and typically constitues less than a few percent of the SM value $C'_7(\mu_b)^\text{SM} = -0.353$. On the other hand, $C'_7$ is suppressed by $m_s/m_b$ in the SM, so the unsuppressed contribution from RS dominates, though its value is still typically smaller than $C_7(\mu_b)^\text{SM}$. 

\begin{figure}
\begin{minipage}{7.4cm}
\includegraphics[width=\textwidth]{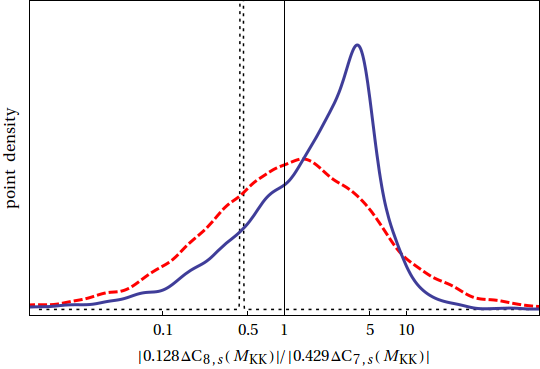}
\end{minipage}
\hfill
\begin{minipage}{7.4cm}
\includegraphics[width=\textwidth]{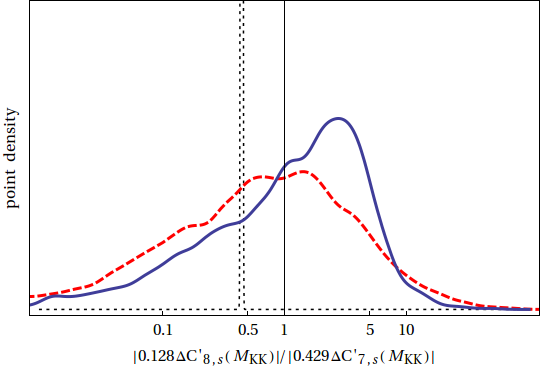}
\end{minipage}
\caption{
Relative sizes of anarchic contributions to the Wilson coefficients $C_7(\mu_b)$ (left) and $C'_{7}(\mu_b)$ (right) from the RG evolution and operator mixing of $\Delta C_8^{(')}$ from $M_\text{KK}$ to $\mu_b$, normalized to the Higgs penguin contribution to $\Delta C_7^{(\prime)}(M_\text{KK})$, with relevant RG evolution factors included. The black (dotted) peak shows the ratio of the Higgs penguin contribution to $\Delta C_8^{(\prime)}(M_\text{KK})$. The red (dashed) and blue (solid) distributions show the ratio of the gluon penguin to $\Delta C_8^{(\prime)}(M_\text{KK})$ for the minimal and custodial model respectively. 
%
\label{fig:C7s_mb_parts}}
\end{figure}

Next, we examine the relative importance of the various RS contributions to the effective $b\to s \gamma$ Wilson coefficients at the  scale $\mu_b$. Fig.\ \ref{fig:C7s_mb_parts} shows the size of the two main anarchic contributions to $\Delta C^{(\prime)}_8(M_\text{KK})$ (see Fig.\ \ref{fig:C8:dominant}a for the relevant Feynman diagrams) normalized to the anarchic contribution to $\Delta C^{(\prime)}_7(M_\text{KK})$ (see Fig.\ \ref{fig:C7:dominant}a). For a straightforward comparison, we also include the relevant RG evolution factors from eq.\ \eqref{eq:RGrunning}.
The ratio of the Higgs penguin contribution to $\Delta C^{(\prime)}_7(M_\text{KK})$ and $\Delta C^{(\prime)}_8(M_\text{KK})$, shown by the black (dotted) peak, is constant and equal for both the minimal and custodial model. As the relevant diagrams depend on the same loop integral and the same combination of Yukawa couplings, their relative size at the KK scale is simply given by the electric charge $Q_u$ of the up-type quark coupled to the photon. After including the RG running down to the scale $\mu_b$, the Higgs penguin contribution to $C_8^{(\prime)}$ turns out to be roughly a 50\% correction to the effect of the anarchic $\Delta C_7^{(\prime)}(M_\text{KK})$ contribution.

The effect of the gluon penguin diagram in $\Delta C^{(\prime)}_8(M_\text{KK})$ depends on a different loop integral and a different combination of Yukawa couplings than the Higgs diagram in $\Delta C^{(\prime)}_7(M_\text{KK})$. Consequently its relative size, again including the relevant RG factors, varies considerably within the minimal (shown in red, dashed line) and the custodial (shown in blue, solid line) model. Observe that the distribution for the minimal model is rather symmetric and peaked around 1, implying that the RS $b\to s\,g$ loop generally contributes as much as the RS $b\to s\gamma$ loop in low energy observables, even yielding the dominant RS contribution in parts of the parameter space. This is in contrast with the SM case, where the $C_8$ contribution only gives a few percent correction to the dominant $C_7$ contribution. In the custodial model the gluon penguin contribution becomes even more important, so that the peak of the distribution gets shifted above 1. Since, as opposed to the Higgs penguin, the additional custodial gluon penguin diagram shown in Fig.\ \ref{fig:custodial} carries the same Yukawa spurion as the minimal model diagram, they simply add constructively, further enhancing the effect of the gluonic penguin contribution. Neglecting these contributions or even the  $C_8^{(\prime)}$ contribution as a whole, as sometimes done in the literature, would therefore be a rather poor approximation. Note that the relative importance of the gluon penguin diagrams depends crucially on the matching of the 5D to the 4D strong gauge coupling. Invoking one loop level matching rather than tree level matching as done here whould reduce their relative size by roughly a factor of four. On the other hand the presence of brane kinetic terms could further enhance the gluonic penguin contribution.


\begin{figure}
\begin{minipage}{7.4cm}
\includegraphics[width=\textwidth]{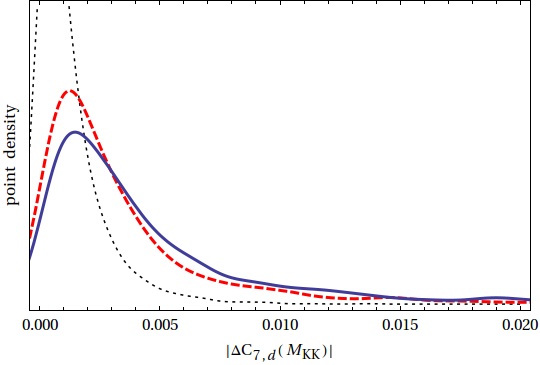}
\end{minipage}
\hfill
\begin{minipage}{7.4cm}
\includegraphics[width=\textwidth]{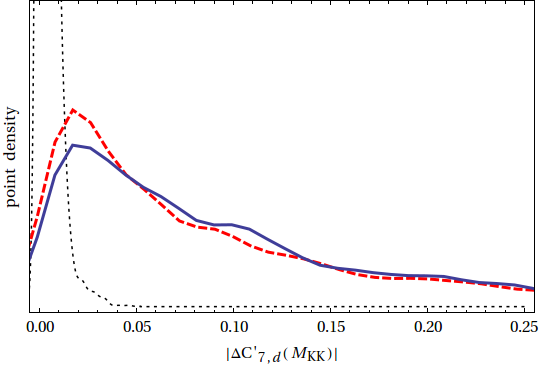}
\end{minipage}\vspace{2mm}

\begin{minipage}{7.4cm}
\includegraphics[width=\textwidth]{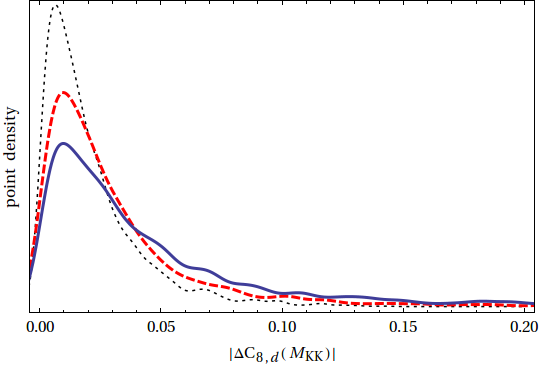}
\end{minipage}
\hfill
\begin{minipage}{7.4cm}
\includegraphics[width=\textwidth]{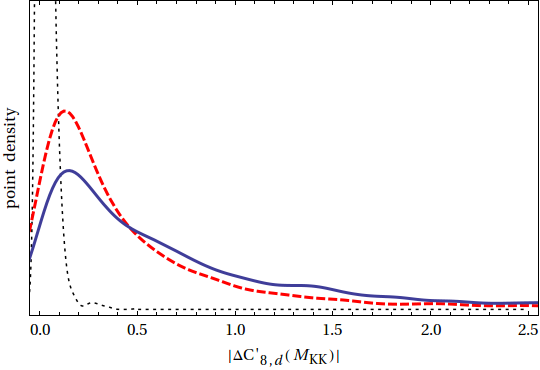}
\end{minipage}
\caption{RS contributions to the $b\to d$ Wilson coefficients $C_7(M_\text{KK})$ (upper left), $C'_{7}(M_\text{KK})$ (upper right), $C_8(M_\text{KK})$ (lower left) and  $C'_{8}(M_\text{KK})$ (lower right)
 in the minimal model (red, dashed), the custodial model (blue, solid), and from the misalignment contribution alone (black, dotted). \label{fig:Cd}}
\end{figure}

For the sake of completeness Fig.\ \ref{fig:Cd} shows the Wilson coefficients for the $b\to d$ system, in analogy to Fig.\ \ref{fig:Cs}. The pattern of effects is very similar to the case of the $b\to s$ system discussed above.

\begin{figure}
\begin{minipage}{7.4cm}
\includegraphics[width=\textwidth]{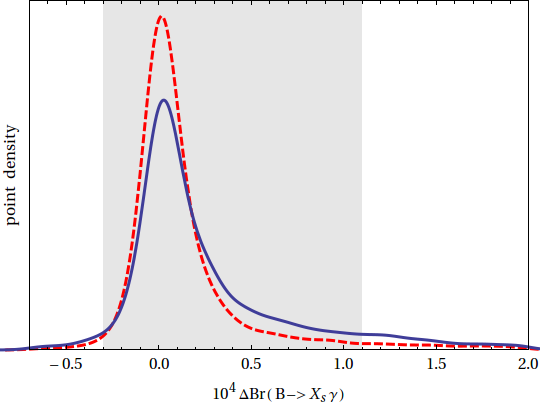}
\end{minipage}
\hfill
\begin{minipage}{7.4cm}
\includegraphics[width=\textwidth]{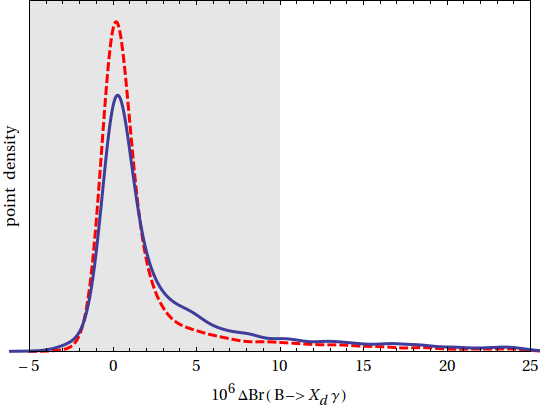}
\end{minipage}
\caption{RS contribution to $\text{Br}(B\to X_s\gamma)$ (left) and $\langle\text{Br}(B\to X_d\gamma)\rangle$ in the minimal (red, dashed) and custodial (blue, solid) model. The experimental constraints according to \eqref{eq:bsg-const} and \eqref{eq:bdg-const} are displayed as grey bands.\label{fig:bsdg-const}}
\end{figure}

Fig.~\ref{fig:bsdg-const} shows the predicted deviations from the SM in the $B\to X_{s,d}\gamma$ branching ratios in the minimal and custodial models. We observe that in both models these branching ratios typically obtain a moderate positive NP contribution well within the current experimental and theoretical uncertainties. Nevertheless, the decays in question put nontrivial constraints on parts of the RS parameter space and should be included in a complete analysis of RS flavor phenomenology. As expected from the size of the Wilson coefficients, the custodial model induces somewhat larger effects than the minimal model.

Interestingly, this pattern of effects is very different from that of the ADD model of a  universal extra dimension \cite{Appelquist:2000nn}, where the KK excitations affect mainly the Wilson coefficient $C_7$, while the opposite-chirality Wilson coefficient $C'_7$ remains very small \cite{Agashe:2001xt,Buras:2003mk}. Since the ADD contribution interferes destructively with the SM contribution, a rather pronounced suppression of $\text{Br}(B\to X_s\gamma)$ is predicted, which was used in \cite{Haisch:2007vb} to derive the bound $1/R > 600 \,\text{GeV}$ on the radius $R$ of the extra dimension.

We also investigated the dependence of the size of the RS contribution to $\text{Br}(B\to X_s\gamma)$ and $\langle\text{Br}(B\to X_d\gamma)\rangle$ on the average Yukawa coupling $Y_*$, but did not find any significant correlation within our parameter scan. These findings at first sight seem to contradict the analytic estimate in section \ref{sec:analytic:estimates:bsgam}. 
Recall that these estimates have been performed in the fully anarchic limit where $Y_*$ is the only free parameter in the Yukawa sector. On the other hand, our scan varies all independent parameters in the flavor sector, so that $\mathcal{O}(1)$ deviations from the fully anarchic ansatz are intrinsic. The dependence on these additional parameters fully hides the dependence on $Y_*$; note also that the latter only varies over an $\mathcal{O}(1)$ range.


\subsection{Effects on benchmark observables}

We now restrict our attention to the custodial model and consider only parameter points that agree with the existing constraints from $\Delta F =2$ transitions, as analyzed in \cite{Blanke:2008zb}. We also impose the bounds from the $B\to X_{s,d}\gamma$ decays as approximated in (\ref{eq:bsg-const}--\ref{eq:bdg-const}), so that all points displayed in the plots lie within the experimentally allowed region.

Since the dipole operators depend on a different combination of RS flavor parameters from the tree level contributions to $\Delta F = 2$ processes \cite{Blanke:2008zb} and $\Delta F = 1$ rare decays \cite{Blanke:2008yr}, observables related to the various sectors are essentially uncorrelated; hence we do not show any numerical results here.

\begin{figure}
\begin{center}
\includegraphics[width=.5\textwidth]{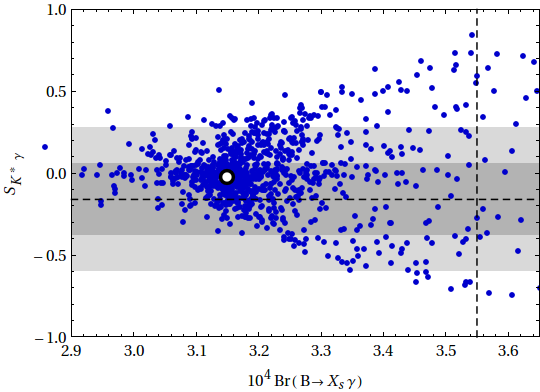}
\end{center}
\caption{CP asymmetry $S_{K^*\gamma}$ as a function of $\text{Br}(B\to X_s\gamma)$. The black-and-white dot indicates the central SM prediction, while the dashed lines show the experimental central values. The grey bands display the experimental $1\sigma$ and $2\sigma$ ranges for $S_{K^*\gamma}$. We omit showing the uncertainty in $\text{Br}(B\to X_s\gamma)$ as it covers the whole range.\label{fig:bsg-SKg}}
\end{figure}

Fig.~\ref{fig:bsg-SKg} shows the correlation between the time-dependent CP asymmetry $S_{K^*\gamma}$ and the branching ratio of $B\to X_s\gamma$. Observe that $S_{K^*\gamma}$ can receive large enhancements relative to its tiny SM value. While non-standard effects in $S_{K^*\gamma}$ are possible for any value of $\text{Br}(B\to X_s\gamma)$, large effects are more likely with enhanced values of the branching ratio. This is related to the fact that RS contributions dominantly affect $C'_{7}$.
While the SM prediction for $B\to X_s\gamma$ is in good agreement with data, it lies below the central value, and an enhancement of this branching ratio is preferred. 
%
%
One can also see that large enhancements are possible in $S_{K^*\gamma}$, and that the present experimental $2\sigma$ range excludes only a small fraction of the RS parameter space.

\begin{figure}
\begin{center}
\includegraphics[width=.5\textwidth]{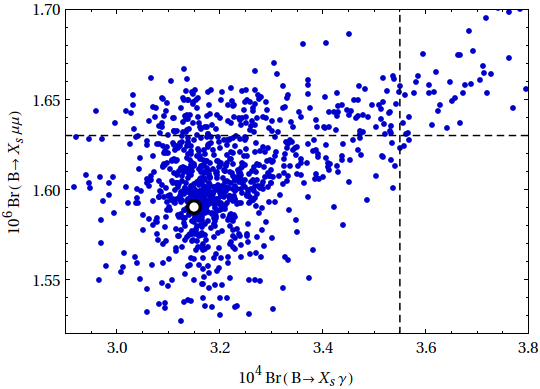}
\end{center}
\caption{Correlation between $\text{Br}(B\to X_s\gamma)$ and $\text{Br}(B\to X_s\mu^+\mu^-)$ for $q^2 \in [1,6] \,\text{GeV}^2$. The black-and-white dot indicates the central SM prediction, while the dashed lines show the experimental central values. We omit showing the experimental and theoretical uncertainties as they cover the whole range.\label{fig:bsg-bsmm}}
\end{figure}

The decay $B\to X_s\mu^+\mu^-$ poses strong constraints on various extensions of the SM, hence it is worth studying it in the custodial RS model. Fig.~\ref{fig:bsg-bsmm} shows the custodial RS branching ratio $\text{Br}(B\to X_s\mu^+\mu^-)$ in the low $q^2$ region as a function of $\text{Br}(B\to X_s\gamma)$. We observe that the enhancement in the custodial RS model is rather small, typically below 10\%. Due to the experimental and theoretical uncertainties involved, this channel does not put any significant constraint on the model. 

\begin{figure}
\begin{center}
\includegraphics[width=.5\textwidth]{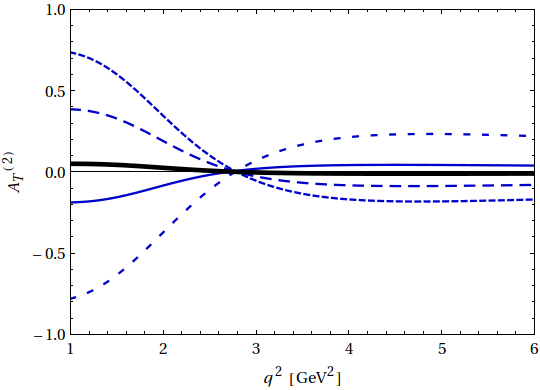}
\end{center}
\caption{Transverse asymmetry $A_T^{(2)}$ as a function of $q^2$, for a few parameter points. The SM prediction is indicated by the thick black line, while each blue line corresponds to an RS parameter point.  \label{fig:AT2q2}}
\end{figure}

Observables far more sensitive to NP in $C'_{7}$ can be constructed from the angular distribution of $B \to K^* \mu^+\mu^-$. Of particular interest is the transverse asymmetry $A_T^{(2)}$, whose $q^2$ dependence is shown in Fig.~\ref{fig:AT2q2}. Observe that large enhancements relative to the small SM value are possible, in particular in the very small $q^2$ region $< 2\,\text{GeV}^2$.
This pattern can be understood from \eqref{eq:AT2}: the $C'_{7}$ contribution  is enhanced at small $q^2$ due to a $1/q^2$ factor, see also \cite{Egede:2010zc,Becirevic:2011bp}. Also note that the custodial RS model predicts a zero crossing for $A_T^{(2)}$ at $q^2\sim 2.7\,\text{GeV}^2$.
%
%
The differential asymmetry would exhibit a very different shape if the dominant NP contribution appeared in $C'_{10A}$. This underlines the model-discriminating power of the $A_T^{(2)}$ asymmetry---in the custodial RS model a deviation from the SM is most likely to be observed for small $q^2$, whereas other models that dominantly affect $C'_{10A}$ predict large effects for larger $q^2$. This pattern is particularly interesting in light of LHCb and the next generation $B$ factories, which will soon be able to measure this asymmetry.

\begin{figure}
\begin{center}
\includegraphics[width=.5\textwidth]{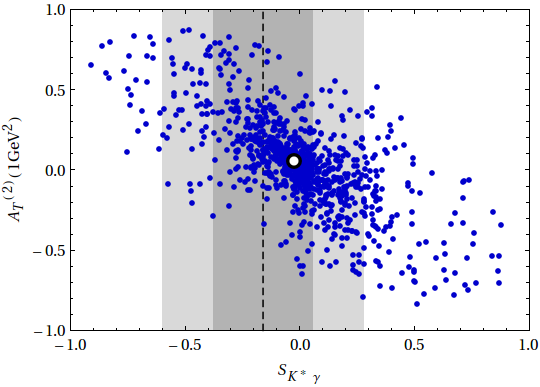}
\end{center}
\caption{Correlation between $S_{K^*\gamma}$ and $A_T^{(2)}(q^2=1\,\text{GeV}^2)$.  The black-and-white dot indicates the central SM prediction, while the dashed line shows the experimental central value. The grey bands display the experimental $1\sigma$ and $2\sigma$ ranges for $S_{K^*\gamma}$.\label{fig:SKg-AT2}}
\end{figure}

Finally, one may consider a possible correlation between $S_{K^*\gamma}$ and $A_T^{(2)}$. Both observables are mostly affected by a large $C'_{7}$, hence some nontrivial correlation can be expected. On the other hand, $S_{K^*\gamma}$ is CP violating while $A_T^{(2)}$ is CP conserving, so the phase of $C'_{7}$ can wash out such correlations. Fig.~\ref{fig:SKg-AT2} shows $A_T^{(2)}(q^2=1\,\text{GeV}^2)$ as a function of $S_{K^*\gamma}$, where a nontrivial linear anti-correlation is seen between the two observables in question. However, this correlation is visibly weakened by the impact of the phase of $C'_{7}$, as expected.



\section{Conclusions}

In this paper we have performed an explicit 5D calculation of the dominant contributions to the Wilson coefficients  $C_7$, $C'_{7}$, and $C_8$, $C'_{8}$ that mediate the $b\to s,d\gamma$ and $b\to s,d\,g$ transitions respectively, in the RS setup with bulk fermions and gauge bosons and an IR-brane localized Higgs. We have evaluated the relevant diagrams for both the minimal scenario with only the SM gauge group in the bulk, and for the custodial model with the electroweak gauge group extended by $\text{SU}(2)_\text{R}$ and a discrete $P_{LR}$ symmetry.
Our main findings from this analysis can be summarized as follows:
\begin{itemize}
\item
The RS contributions to $C'_7$ typically exceed those to $C_7$ by an order of magnitude, and the latter remain a rather small correction to the SM value. This pattern can be understood by considering the bulk profiles of the quark fields involved: the primed Wilson coefficient describes the decay of a left-handed $b$ quark, which, being localized towards the IR brane, is more sensitive to flavor violating effects than the right-handed $b$ quark entering $C_7$. Analogous comments apply regarding the hierarchy $C_8 \ll C'_8$.
\item
Contrary to the SM, where $C_8 < C_7$, RS contributions to the gluonic penguins are larger than the ones to the photonic penguins. This results from the large contributions from the diagram containing the non-abelian triple gluon (KK gluon) vertex, which is absent in $C^{(\prime)}_7$ and does not change flavor in the SM. In addition, the renormalization group mixing of $C^{(\prime)}_7$ and $C^{(\prime)}_8$ is more pronounced due to the large separation of the $M_\text{KK}$ and $m_b$ scales. Consequently, gluonic penguin contributions have a significant impact on $b\to s,d\gamma$, comparable to or larger than the photonic penguin contribution. This is in contrast to the SM,  where they yield only a few percent correction to the photonic Wilson coefficients at the $m_b$ scale. 
\item
In all cases, the dominant effect comes from the anarchic contributions, which are not aligned with the SM quark mass matrices.
However, the unprimed (right to left) operators pick up appreciable contributions from misalignment diagrams, which are proportional to the SM quark mass matrices up to a dependence on the bulk spectrum. This is because, in contrast to the anarchic diagrams, the misalignment diagrams are not suppressed by the $b_R$ wave function relative to the $b_L$ wavefunction, as explained in Appendix~\ref{sec:misalignment}. 
\item
The impact  on the Wilson coefficients in question is somewhat larger in the  custodial model than in the minimal model, since the extended fermion content that was introduced to reconcile the model with the $Zb\bar b$ constraint yields additional contributions.
\end{itemize}

For a study of the phenomenological implications of these new contributions, we restricted our attention to the custodial model since the minimal model is not consistent with electroweak precision constraints for low KK masses $M_\text{KK} = 2.5\,\text{TeV}$. To this end, following \cite{Blanke:2008zb} we performed a parameter scan of the 5D bulk masses and fundamental Yukawa coupling matrices, imposing constraints from quark masses and CKM parameters and from meson-antimeson mixing. We studied the bounds provided by the branching ratios $\text{Br}(B\to X_s\gamma)$ and   $\langle\text{Br}(B\to X_d\gamma)\rangle$ and the effects in a number of benchmark observables, namely the time-dependent CP asymmetry $S_{K^*\gamma}$, the inclusive branching ratio $\text{Br}(B\to X_s\mu^+\mu^-)$ and the forward-backward asymmetry $A_\text{FB}$ and the transverse asymmetry $A_T^{(2)}$ in $B\to K^*\mu^+\mu^-$, where we found the following patterns:
\begin{itemize}
\item
The branching ratios of the radiative inclusive $B\to X_{s,d}\gamma$ decays provide a non-negligible constraint on RS models and exclude roughly 15\%  of the parameter points generated for the custodial model that were in agreement with bounds from $\Delta F=2$ observables. A complete phenomenological study should therefore take these constraints into account. However, since the major part of parameter space survives, no useful bound on the KK scale can be derived.
\item
Due to more precise data and SM theory prediction, $\text{Br}(B\to X_s\gamma)$ generally puts a stronger constraint on the RS parameter space than $\langle\text{Br}(B\to X_d\gamma)\rangle$.
The latter observable is still useful as it yields complementary information on the allowed parameter space.
\item
As the RS contributions enter dominantly through the primed operators, a modest enhancement of the $B\to X_{s,d}\gamma$ branching ratios can be expected, although a slight suppression is not rigorously excluded. Such an enhancement would be welcome in $B\to X_s\gamma$, where the data lie somewhat above the SM value, albeit still in good agreement. On the other hand, for $B\to X_d\gamma$ the central values of the SM and the data are in excellent agreement and the uncertainties are sizable, and no prefered sign for the NP contribution can be deduced.
\item
The inclusive branching ratio $\text{Br}(B\to X_s\mu^+\mu^-)$ and the forward backward asymmetry $A_\text{FB}$ in $B\to K^*\mu^+\mu^-$ receive very small corrections from RS physics and remain in good agreement with recent data. While we restricted our analysis to the low $q^2$ region, these statements also apply to the high $q^2$ region since the latter region is mostly sensitive to NP in the electroweak Wilson coefficients $C^{(\prime)}_{9V,10A}$, which remain SM-like in the custodial model.
\item
We identify the time-dependent CP asymmetry $S_{K^*\gamma}$ in $B\to K^*\gamma$ decays and the transverse asymmetry $A_T^{(2)}$ in the low $q^2$ region of $B\to K^*\mu^+\mu^-$ as promising benchmark observables to look for large effects generated by the custodial RS model. Both observables are known to be very sensitive to the primed Wilson coefficients, in particular $C'_7$, which is dominantly affected by RS contributions. Furthermore, studying the $q^2$ dependence of $A_T^{(2)}$ allows for a clear distinction of models such as the custodial RS model that dominantly affect $C'_7$ from models that predict large NP effects in the electroweak Wilson coefficient $C'_{10A}$.  
\end{itemize}

In summary, our analysis shows that radiative and semileptonic $B$ decays offer intriguing possibilities to find deviations from the SM generated by RS KK modes and anarchic Yukawa structure. If such effects are found at the LHCb and the next generation $B$ factories, it will be particularly interesting to study the plethora of observables provided by these decay modes in a correlated manner, which offers the ability to distinguish RS with custodial symmetry from other NP scenarios that predict a different pattern of effects.

\subsection*{Acknowledgments}
We especially thank Kaustubh Agashe for many extensive discussions.
We further thank  
Wolfgang Altmannshofer,
Christoph Bobeth,
	Andrzej Buras, 
	Csaba Cs\'aki,
	Stefania Gori,
	Yuval Grossman,
Gudrun Hiller,
	Jay Hubisz,
	Enrico Lunghi,
	Paride Paradisi,
	Gilad Perez,
        Maxim Perelstein,
David Straub,
        and
	Andreas Weiler
for discussions and useful comments. 
We also thank
	Kaustubh Agashe,
	Martin Beneke,
	Andrzej Buras,
and
	Csaba Cs\'aki
for valuable comments on the manuscript.
The authors are supported in part by the U.S. National Science Foundation through grant PHY-0757868 and CAREER award PHY-0844667.
P.T.\ is also supported by a Paul \& Daisy Soros Fellowship for New Americans. 
Y.T.\ is supported by a Fermilab Fellowship in Theoretical Physics. Fermilab is operated by Fermi Research Alliance, LLC, under Contract DE-AC02-07CH11359 with the United States Department of Energy. 
M.B.\ acknowledges partial support by the National Science Foundation under Grant No. 1066293 and the hospitality of the Aspen Center for Physics. 
P.T.\ and Y.T.\ would like to thank the Kavli Institute for Theoretical Physics (KITP) where part of this work was completed. The KITP is supported in part by the National Science Foundation under Grant No. NSF PHY05-51164.
B.S. acknowledges the hospitality of the Theoretical Advanced Study Institute (TASI-11) at the University of Colorado at Boulder, where part of this work was completed. 

\appendix

\section{Dimensionless Integrals for Leading Diagrams}\label{app:integrals}

This appendix defines the dimensionless integrals associated with the leading contributions to the $a$ and $b$ terms of the dipole Wilson coefficients $C_{7,8}$ in Section \ref{sec:penguins}. Details of the derivation of these integrals are found in the appendix of \cite{Csaki:2010aj}. In the mass insertion approximation the Standard Model contribution appears as an infrared pole, which we subtract.

\subsection{Propagator functions}
\label{subsec:propagators}

We use dimensionless integration variables $x \equiv k_E z \in [wy,y]$ and $y\equiv k_E R'\in [0,\infty]$, where $k_E$ is the Euclidean loop momentum and $w=(R/R')$ is the warp factor. 
The integrals are expressed with respect to the functions that appear in the mixed position--Euclidean momentum space fermion propagator,
\begin{align}
    \Delta(k_E,x,x') &\equiv i\frac{R'}{w^4}\bar{\mathcal D}\tilde F_y^{xx'} =
    \begin{pmatrix}
    y\tilde D_-\tilde F_- & \sigma^\mu y_\mu \tilde F_+ \\ 
    \bar\sigma^\mu y_\mu \tilde F_- & y\tilde D_+ \tilde F_+
    \end{pmatrix}, \quad\quad\quad \tilde D_\pm \equiv \pm\left(\partial_x-\frac 2x\right)+\frac cx.\label{eq:warped:G:fromF:Euclidean}
\end{align}
where the $\tilde F$ functions are defined for $x>x'$ (i.e.\ $z>z'$) by
\begin{align}
\tilde F^L_{-} &=\phantom{+} \frac{(xx')^{5/2}}{y^5} \frac{S_{c_L}(x_-,y_-) S_{c_L}(x'_-,wy_-)}{S_{c_L}(y_-,wy_-)}
&
\tilde F^L_{+} &=- \frac{(xx')^{5/2}}{y^5} \frac{T_{c_L}(x_+,y_-) T_{c_L}(x'_+,wy_-)}{S_{c_L}(y_-,wy_-)}
\label{eq:tilde:F:1}
\\
\tilde F^R_{-} &=- \frac{(xx')^{5/2}}{y^5} \frac{T_{c_R}(x_-,y_+) T_{c_R}(x'_-,wy_+)}{S_{c_R}(y_+,wy_+)}
&
\tilde F^R_{+} &=\phantom{+} \frac{(xx')^{5/2}}{y^5} \frac{S_{c_R}(x_+,y_+) S_{c_R}(x'_+,wy_+)}{S_{c_R}(y_+,wy_+)}.
\label{eq:tilde:F:2}
\end{align}
The analogous functions for $x<x'$ are given by replacing $x\leftrightarrow x'$ in the above formulas. $S$ and $T$ function are products of Bessel functions,
\begin{align}
	S_c(x_\pm,x'_\pm) &= I_{c\pm 1/2}(x) K_{c\pm 1/2}(x') - I_{c\pm 1/2}(x') K_{c\pm 1/2}(x)\\
	S_c(x_\pm,x'_\mp) &= I_{c\pm 1/2}(x) K_{c\mp 1/2}(x') - I_{c\mp 1/2}(x') K_{c\pm 1/2}(x)\\
	T_c(x_\pm,x'_\mp) &= I_{c\pm 1/2}(x) K_{c\mp 1/2}(x') + I_{c\mp 1/2}(x') K_{c\pm 1/2}(x).
\end{align}
Similarly, the mixed position--Euclidean momentum space vector propagators are $-i\eta^{\mu\nu} G$ and $i\bar G$ for the 4-vector and scalar parts respectively. For $x<x'$, the $G$ functions are,
\begin{align}
	G_k(z,z') &= \frac{(R')^2}{R} G_y(x,x') = \frac{(R')^2}{R} \frac{xx'}{y} \frac{T_{10}(x,y)T_{10}(x',wy)}{S_{00}(wy,y)},\label{eq:G:def}\\
	G_{5k}(z,z') &= \frac{(R')^2}{R} \bar G_y(x,x') = \frac{(R')^2}{R} \frac{xx'}{y} \frac{S_{00}(x,y)S_{00}(x',wy)}{S_{00}(wy,y)},\label{eq:Gbar:def}
\end{align}
where 
\begin{align}
	T_{ij}(x,y) &= I_i(x)K_j(y)+ I_j(y)K_i(x)\\
	S_{ij}(x,y) &= I_i(x)K_j(y) - I_j(y)K_i(x).
\end{align}
For $z<z'$ the above formula is modified by $x\leftrightarrow x'$.

\subsection[$C_7$ integrals]{\boldmath $C_7$ integrals}

We label vertices such that the external fermion legs attach to vertices 1 and 3, and the photon or gluon is emitted at vertex 2. Propagators attached to the brane $x=y$ signify Yukawa couplings or mass insertions, which may change the fermion flavor as labeled by its bulk mass, $c$. We have left this $c$ dependence implicit in the following expressions.

\begin{align}
	I_{C_{7a}} =&\int dy \, dx\, y^2 \left(\frac yx\right)^4\Big[ 
	-2 \tF{+}{L}{y}{y}{x}\,
	\tF{+}{L}{y}{x}{y}\,
	\tF{-}{R}{y}{y}{y}
	\frac{y^2}{y^2+(M_WR')^2}
	\nonumber\\
	&
	+\tF{+}{L}{y}{y}{x}\,
	\tF+Lyxy \,
	\tF-Ryyy
	\frac{y^4}{(y^2+(M_WR')^2)^2}
	-
	\frac 12 \left(y\,\partial_{k_E}\tF+L{y}yx\right)
	\tF+Lyxy \,
	\tF-Ryyy
	\frac{y^2}{y^2+(M_WR')^2}
	\nonumber\\
	&-
	\frac 12 \left(y\,\partial_{k_E} \tDF-L{y}yx\right)
	\tDF+Lyxy \,
	\tF-Ryyy\,
	\frac{1}{y^2+(M_WR')^2}
	+2
	\tF+Lyyy\,
	\tDF+Ryyx\,
	\tDF-Ryxy
	\frac{1}{y^2+(M_WR')^2}
	\nonumber\\
	&
	-\tF+Lyyy \,
	\tDF+Ryyx \,
	\tDF-Ryxy
	\frac{y^2}{(y^2+(M_WR')^2)^2}
	+
	\frac 12 \left(y\,\partial_{k_E} \tF+L{y}yy\right)
	\tDF+Ryyx\,
	\tDF-Ryxy
	\frac{1}{y^2+(M_WR')^2}
	\nonumber \\
	&+
	\tF+Lyyy\,
	\tF-Ryyx\,
	\tF-Ryxy
	\frac{y^2}{y^2+(M_WR')^2}
	+
	\frac 12 \left(y\,\partial_{k_E} \tF+L{y}yy\right)
	\tF-Ryyx \,
	\tF-Ryxy
	\frac{y^2}{y^2+(M_WR')^2}
	\nonumber \\
	&
	+ \frac 12 \tF+Lyyy\,
	\left(y\,\partial_{k_E} \tF-R{y}yy\right)
	\tF-Ryxy
	\frac{y^2}{y^2+(M_WR')^2}
	\nonumber \\
	&
	+
	\frac 12 \tF+Lyyy\,
	\left(y\,\partial_{k_E} \tDF+R{y}yx\right)
	\tDF-R{y}xy 
	\frac{1}{y^2+(M_WR')^2}
	\Big].\label{eq:I:1MIH0}
	\end{align}
	The $C_{7b}$ integral is the sum of two parts corresponding to diagrams with an internal gluon ($G$) or scalar gluon ($G_5$) in the loop,
\begin{align}
	I_{C_{7b}} =&\, I_{C_{7b}}^{(G)} + I_{C_{7b}}^{(G_5)}\label{eq:I:c7b}.
\end{align}
Each of these terms include diagrams with a single mass insertion, either on the incoming, internal, or outgoing fermion line. 
\begin{align}
	I_{C_{7b}}=&\int dy\, dx_1\, dx_2\, dx_3\,y\,\left(\frac{y}{x_2}\right)^{4}\partial_{k_E}G^{31}\,\nonumber\\
&\Big\{\frac{1}{2}\,
		\left(\frac{y}{x_1}\right)^{2+c_L}
		\left(\frac{y}{x_3}\right)^{4}\,\tDF{+}{L}{(m_bR')}{(x_3 m_bR'/y)}{(m_bR')}
		\left(\tDF{-}{L}{y}{1}{2}\,\tF{-}{L}{y}{2}{3}+\tF{+}{L}{y}{1}{2}\,\tDF{-}{L}{y}{2}{3}\right)\nonumber\\
		&+\frac{1}{2}\,
		\left(\frac{y}{x_1}\right)^{4}
		\left(\frac{y}{x_3}\right)^{2-c_R} \tDF{+}{R}{(m_bR')}{(m_bR')}{(x_1 m_bR'/y)}
		\left(\tDF{-}{R}{y}{1}{2}\,\tF{-}{R}{y}{2}{3}+\tF{+}{R}{y}{1}{2}\,\tDF{-}{R}{y}{2}{3}\right) \nonumber\\
		&+\left(\frac{y}{x_1}\right)^{2+c_L}
		\left(\frac{y}{x_3}\right)^{2-c_R}
		\Big(-\tDF{+}{R}{y}{3}{2}\,\tDF{-}{R}{y}{2}{y}\,\tF{+}{L}{y}{y}{1}\,+y^2\,\tF{-}{R}{y}{3}{2}\,\tF{-}{R}{y}{2}{y}\,\tF{+}{L}{y}{y}{1}\,\nonumber\\
		&-\tDF{-}{L}{y}{y}{2}\,\tDF{+}{L}{y}{2}{1}\,\tF{-}{R}{y}{3}{y}\,+y^2\,\tF{-}{R}{y}{3}{y}\,\tF{+}{L}{y}{y}{2}\,\tF{+}{L}{y}{2}{1}\, \Big)
		\Big\}
\end{align}
\begin{align}
I_{C_{7b}}'=&\int dy\, dx_1\, dx_2\, dx_3\,\frac{1}{2}\,\left(\frac{y}{x_2}\right)^{4}\nonumber\\
&\Big\{
		\left(\frac{y}{x_1}\right)^{2+c_L}
		\left(\frac{y}{x_3}\right)^{4}\,\tDF{+}{L}{(m_bR')}{(x_3 m_bR'/y)}{(m_bR')}\times\nonumber\\
		&\left(\tF{-}{L}{y}{1}{2}\,\tDF{+}{L}{y}{2}{3}\,(y\,\partial_{k_E}G_5^{31}+4\,G_5^{31})+y\,G_5^{31}(\tDF{+}{L}{y}{2}{3}\,\partial_{k_E}\tF{-}{L}{y}{1}{2}\,-\tF{+}{L}{y}{2}{3}\,\partial_{k_E}\tDF{+}{L}{y}{1}{2}\,)\right)\nonumber\\
		&+\left(\frac{y}{x_1}\right)^{4}
		\left(\frac{y}{x_3}\right)^{2-c_R} \tDF{+}{R}{(m_bR')}{(m_bR')}{(x_1 m_bR'/y)}\times\nonumber\\
		&\left(\tF{-}{R}{y}{1}{2}\,\tDF{+}{R}{y}{2}{3}\,(y\,\partial_{k_E}G_5^{31}+4\,G_5^{31})+y\,G_5^{31}(\tDF{+}{R}{y}{2}{3}\,\partial_{k_E}\tF{-}{R}{y}{1}{2}\,-\tF{+}{R}{y}{2}{3}\,\partial_{k_E}\tDF{+}{R}{y}{1}{2}\,)\right)\nonumber\\
		&+\left(\frac{y}{x_1}\right)^{2+c_L}
		\left(\frac{y}{x_3}\right)^{2-c_R}\times\nonumber\\
		&\Big(\tDF{+}{L}{y}{1}{2}\,(4\,+y\,\partial_{k_E})(\tF{+}{L}{y}{2}{y}\,\tDF{+}{R}{y}{y}{3}\,G_5^{13})-y\,\tF{-}{L}{y}{1}{2}\partial_{k_E}\,(\tDF{+}{L}{y}{2}{y}\,\tDF{+}{R}{y}{y}{3}\,)\,G_5^{13}\nonumber\\
		&+\tDF{+}{L}{y}{1}{y}\,\tDF{+}{R}{y}{y}{2}\,(4\,+y\,\partial_{k_E})(\tF{+}{R}{y}{2}{3}\,G_5^{13})\,-y\,\tDF{+}{L}{y}{1}{y}\,\tF{-}{R}{y}{y}{2}\,G_5^{13}\,\partial_{k_E}\,\tDF{+}{R}{y}{2}{3}\,\Big)
		\Big\}
		\end{align}

\subsection[$C_8$ integrals]{\boldmath $C_8$ integrals}

The $C_{8a}$ integral contains a piece identical to the $C_{7a}$ integral associated with the charged Higgs loop as well as gluon loop diagrams with three mass insertions,
\begin{align}
	I_{C_{8a}} =&\,  I_{C_{8a}}^{(1)} + 2 I_{C_{8a}}^{(2)} + I_{C_{8a}}^{(3)}. \label{eq:I:c8a}
\end{align}
The gluon loops are labeled by the number of internal mass insertions, so that $I_{C_{8a}}^{(1)}$ is associated with the diagram with an external mass insertion on each leg, and the factor of two on $I_{C_{8a}}^{(2)}$ accounts for the two possible placements of the external mass insertion\footnote{
These integrands differ by $L\leftrightarrow R$, but the integrals are approximately the same. 
}.
%
%
\begin{align}
		I_{C_{8a}}^{(1)} =& \int dy\, dx_1 dx_2 dx_3 
		\left(\frac{y}{x_1}\right)^{4}
		\left(\frac{y}{x_2}\right)
		\left(\frac{y}{x_3}\right)^{4}\times \nonumber\\
		&\tDF{+}{R}{y_s}{y}{1}\,\tDF{-}{R}{y}{1}{0}\,\tDF{-}{L}{y}{y}{3}\,\tDF{+}{L}{y_b}{3}{y}	\Big\{-\frac{5}{2}y\partial_{k_E}\Big(G^{12}_y\,G^{23}_y\Big)+10\,G^{12}_y\,G^{23}_y\Big\},\label{eq:1p2mig}
		\\
		I_{C_{8a}}^{(2)} =& \int dy\, dx_1 dx_2 dx_3 
		\left(\frac{y}{x_1}\right)^{2+c_L}
		\left(\frac{y}{x_2}\right)
		\left(\frac{y}{x_3}\right)^{4}y^3\,\times\nonumber\\
		&\tF{+}{L}{y}{1}{y}\,\tF{-}{R}{y}{y}{y}\,\tDF{-}{L}{y}{y}{3}\,\tDF{+}{L}{(m_bR')}{(x_3m_bR'/y)}{(m_bR')}\,\partial_{k_E}(G^{12}_y\,G^{23}_y)\label{eq:2mig}	
		\\
		I_{C_{8a}}^{(3)} =& \int dy\, dx_1 dx_2 dx_3  
		\left(\frac{y}{x_1}\right)^{2+c_L}
		\left(\frac{y}{x_2}\right)
		\left(\frac{y}{x_3}\right)^{2-c_R}y^2\,
		\times \nonumber\\
		&\tF{+}{L}{y}{1}{y}\,\tF{-}{R}{y}{y}{3}\,\tF{+}{L}{y}{y}{y}\,\tF{-}{R}{y}{y}{y}\,\Big\{-\frac{5}{2}y\,\partial_{k_E}\Big(G^{12}_y\,G^{23}_y\Big)+10G^{12}_yG^{23}_y\Big\}.\label{eq:3mig}
\end{align}
For $C_{8b}$, the only dominant diagram is the gluon loop with an internal mass insertion. All other analogous diagrams (e.g.\ mass insertion on an external leg, or loops with $G^5$) contain no zero modes and hence give negligible contributions after alignment. 
\begin{align}
	I_{C_{8b}} =&\, 
	\int dy\, dx_1 dx_2 dx_3 
		\left(\frac{y}{x_1}\right)^{2+c_L}
		\left(\frac{y}{x_2}\right)
		\left(\frac{y}{x_3}\right)^{2-c_R}y^2
		\times \nonumber\\
		&\tF{+}{L}{y}{1}{y}\,\tF{-}{R}{y}{y}{3}\,\Big\{-\frac{5}{2}y\,\partial_{k_E}\Big(G^{12}_y\,G^{23}_y\Big)+10\,G^{12}_y\,G^{23}_y\Big\}.\label{eq:I:c8b}
\end{align}

\section{Charged Higgs diagram calculation}
\label{sec:sample:calc} 

As an example of how to calculate diagrams in the mixed position/momentum formalism, we present the calculation of the leading contribution to the anarchic piece of the $C_7$ operator coming from the charged Higgs diagram in Fig.~\ref{fig:C7:a}.
As discussed in Section~\ref{sec:structure:of:amplitude}, it is sufficient to compute the coefficient of the $p_\mu$ term in the amplitude. This allows us to directly write the finite physical contribution to the amplitude without worrying about regularization of potentially divergent terms.  
In addition to the bulk fermion propagators in mixed position/momentum space, $\Delta(p,z,z')$, which are given in Appendix~\ref{subsec:propagators}, the relevant Feynman rules are given by
\begin{center}
 \begin{tikzpicture}[line width=1.5 pt, scale=1]
     \draw [fermion] (-1, 1) -- (0, 0);
     \draw [fermion] (0, 0) -- ( 1, 1);
 	\draw [scalarnoarrow, dash pattern=on 6pt off 3pt] ( 0, 0) -- (0, -1);
     \node at (-1.2, 1.2) {$D$};
     \node at ( 1.2, 1.2) {$Q$};
     \node at (0, -1.3) {$H$}; 
 	\node at (3.0,0){\large$\displaystyle= \left(\frac{R}{R'}\right)^3 Y_5$};
 \end{tikzpicture}
 \quad\quad\quad
 \begin{tikzpicture}[line width=1.5 pt, scale=1]
     \draw [fermion] (-1, 1) -- (0, 0);
     \draw [fermion] (0, 0) -- ( 1, 1);
 	\draw [vector] ( 0, 0) -- (0, -1);
     \node at (-1.2, 1.2) {$f$};
     \node at ( 1.2, 1.2) {$f$}; 
     \node at (0, -1.3) {$A_\mu$}; 
 	\node at (3,0){\large$\displaystyle= \left(\frac{R}{z}\right)^4 e\gamma^\mu$};
 \end{tikzpicture}
\end{center}
A derivation of the propagators and a more complete set of Feynman rules is given the appendix of \cite{Csaki:2010aj}.
The amplitude for the diagram with a $b$ of momentum $p$ decaying into a photon of momentum $-q$ and a $s$ of momentum $p'$ is
\begin{align}
    \mathcal M^\mu &= \frac{ev}{\sqrt{2}}
    \frac{R^8}{R'^6}
    Y_{s k}Y^\dag_{k\ell}Y_{\ell b}
    \int \frac{d^4k}{(2\pi)^4}\int_R^{R'}dz
    \left(\frac Rz\right)^4 
    \bar u_{Q_s}(p') f_{Q_s} \left[G^\mu\right]_{k\ell}  f_{D_{b}}u_{D_b}(p)\,\Delta_H(k-p)\label{eq:amplitude:general}
\end{align} 
where $k$ and $\ell$ index the flavors of the internal fermions and $\Delta_H$ is the 4D Higgs propagator. Writing $k'=k+q$, the Dirac structure $G^\mu$ for the diagram with the mass insertion before (a) or after (b) the photon emisison is
\begin{align}
    \left[G^\mu_{(a)}\right]_{k\ell} &= \Delta_{D_k}(k',R',z)\;\gamma^\mu\;\Delta_{D_k}(k,z,R')\;\Delta_{Q_\ell}(k,R',R'),\\
    \left[G^\mu_{(b)}\right]_{k\ell} &= \Delta_{D_k}(k',R',R')\;\Delta_{Q_\ell}(k,R',z)\;\gamma^\mu\;\Delta_{Q_\ell}(k,z,R').
\end{align}
We may now expand the fermion propagators in terms of scalar  functions $F$, which are the Minkowski space versions of the $\tilde F$ functions defined in Appendix~\ref{subsec:propagators} to simplify the Dirac structure and write the integrand in the form
\begin{align}
    \bar{u}_{Q_s}
    \left(
    \bar{g}^{(n)}_{k\ell}\gamma^\mu\slashed{k}
    +g^{(n)}_{k\ell}\slashed{k'}\gamma^\mu
    \right)
    P_R\,
    u_{D_b}\,\Delta_H(k-p) \quad\quad\quad\quad\quad n\in\left\{a,b\right\},
\end{align}
where $g^{(n)}$ is a scalar function that takes the form
\begin{align}
    g^{(a)}_{k\ell}(z,k,k') &= k^2 \left[F^{-}_{D_k}(k',R',z)\right] \left[ F^{-}_{D_k}(k,z,R')\right]\left[F^{+}_{Q_\ell}(k,R',R')\right]\\
    g^{(b)}_{k\ell}(z,k,k') &= \left[F^{-}_{D_k}(k',R',R')\right] \left[\tilde D_- F^{-}_{Q_\ell}(k',R',z)\right]\left[\tilde D_+F^{+}_{Q_\ell}(k,z,R')\right].
\end{align}
The derivative operators $\tilde D_\pm$ are defined in (\ref{eq:warped:G:fromF:Euclidean}). $\bar g^{(n)}$ has a similar definition but, as we show below, drops out of the final expression.

To identify the $p^\mu$ coefficient, which in turn determines the coefficient of the $C_7$ effective operator, Taylor expand in $p$ and $q$ and perform the integral. It is sufficient to take only the leading order terms since higher terms are suppressed by the ratio of the external fermion masses to the characteristic loop energy scale (e.g.\ $m_H$ or $1/R'$). The terms proportional to $g^{(a)}$ and $g^{(b)}$ thus can be expanded as
\begin{align}
    g(z,k,k') \slashed{k}' \gamma^\mu \Delta_H(k-p) =  \left.\left(g +\frac{\partial g}{\partial k'}\frac{k\cdot q}{k}  \right)\right|_{k'=k} \left(\slashed{k}+\slashed{q}\right) \gamma^\mu \left.\left(\Delta_H(k)+2k\cdot p\,\Delta^2_H(k)  \right)\right.
\end{align}
The $\bar{g}$ terms yield expressions proportional to $\gamma^\mu\slashed{p}$ and $\gamma^\mu\slashed{p'}$. By using the Clifford algebra and the equations of motion for the external particles one can show that these terms are proportional to $ m_b\gamma^\mu$ and $2p'^\mu - m_s\gamma^\mu$ respectively. Thus these terms can be ignored since these do not contribute to the $p^\mu$ coefficient.
The $g$ terms, on the other hand, contribute expressions of the form
\begin{align}
    \left(
    k^2 g\Delta_H^2(k) - \frac 12 k\frac{\partial g}{\partial k'}\Delta_H(k) - 2g\Delta_H(k)
    \right)p^\mu,
\end{align}
where we write $k=\sqrt{k_\mu k^\mu}$ and $g$ is evaluated at $q=0$, i.e.\ $k'=k$. 

Finally, the coefficient $a_{k\ell}$ of the amplitude  (\ref{eq:MasterAmplitude}) can be written with respect to the Wick-rotated integral of the prefactor multiplying $p^\mu$,
\begin{align}
    a_{k\ell} &= -2i \sum_{n=a,b}\int_0^\infty  dy \int_{0}^y dx \; y^2\left(\frac{y}{x}\right)^4\left\{ y^2 g^{(n)} \Delta_H + \frac{y}{2}\frac{\partial g^{(n)}}{\partial y'} + 2 g^{(n)} \right\}\Delta_H,
\end{align}
where we have defined the dimensionless integration variables  $x= -ikz$, $y = -ikR'$, and $y'=-i(k+q)R'$. This is equivalent to replacing the Minkowski space functions $F(kR',zR', z'R')$ with the Euclidean space functions $\tilde F(y,x,x')$ defined in Appendix~\ref{subsec:propagators}.
The $g$ and $\Delta_H$ functions are evaluated at $k\to iy$ and $m_H\to m_HR'$.
These are now completely scalar expressions that can be evaluated numerically. The explicit form of the integrand is given in (\ref{eq:I:1MIH0}). 

Other diagrams are calculated following a similar algorithm with the caveat that diagrams with bulk gauge bosons have 5D propagators, which carry additional space integrals over the extra dimension.

\section{Estimating the size of  the misalignment contribution}
\label{sec:misalignment}

In this appendix we clarify a subtlety in the size of the anarchic contributions ($\Delta C_{7,8a}^{(\prime)}$) versus the misalignment contributions ($\Delta C_{7,8b}^{(\prime)}$) to the Wilson coefficients, as defined in Section~\ref{sec:structure:of:amplitude}.
For the anarchic contributions the relative sizes of the right-to-left (unprimed) coefficients to the left-to-right (primed) coefficients are given by the relative size of the $f_{b_L}$ and $f_{b_R}$ wavefunctions on the IR brane. On the other hand, the misalignment contributions for the two chiral transitions do not follow this pattern and are, in fact, of the same order of magnitude. We show here that this apparent inconsistency can be understood by accounting for cancelations coming from the rotation to the SM fermion mass basis. 

For simplicity, consider the $2\times 2$ matrix of misalignment diagrams $q^R_j \to q^L_i$ where we only consider the second and third generations. This transition is given by the $b_{ij}$ term in (\ref{eq:MasterAmplitude}), which we may parameterize as
\begin{align}
\label{eq:misalignment:estimate}
\text{(misalignment term)}_{ij} &\sim
\begin{pmatrix}
(b-c-d) \, y_{11}		& (b-c+d)\, y_{12}\\
(b+c-d) \, y_{21}		& (b+ c+d)\, y_{22}
\end{pmatrix}.
\end{align}
Here we have written $b$ as an average scale for the $b_{ij}$ matrix, and $y_{ij} = f_{Q_i} Y^\dag_{d\,ij} f_{D_j}$. The $c\sim 10^{-1}$ and $d\sim 10^{-2}$ terms represent deviations from the average. In particular, the $c$ deviations account for the effect of an internal $b_L$ (whose zero mode profile is very different from that of the light quarks)  while the $d$ deviations account for the smaller effect of an internal $b_R$. 

In order to pass to the physical basis, one must apply to this matrix the same rotation that diagonalizes the SM mass matrix, which is proportional to $y$. The off-diagonal terms of the rotated misalignment matrix give the $C_7$ and $C_7'$ coefficients (the argument for $C_8$ is identical),
\begin{align}
\label{eq:misalignment:rotation}
\begin{pmatrix}
1 & \delta \\
\delta' & 1
\end{pmatrix}
\text{(misalignment term)}
\begin{pmatrix}
1 & \gamma \\
\gamma' & 1
\end{pmatrix}
\;\sim\;
\begin{pmatrix}
 & C_{7b} \\
C_{7b}'^\dag & 
\end{pmatrix}.
\end{align}
The parameters $\delta$ and $\gamma$ are ratios of the left- and right-handed zero mode wavefunctions on the brane; the primed and unprimed parameters are related by a minus sign. 

We focus on order of magnitude estimates, so we introduce a numerical parameter $\epsilon \sim 10^{-1}$. Normalizing the Yukawa to $y_{22} = 1$, our parameters are approximately
\begin{align}
c \sim \epsilon
\qquad
d \sim  \epsilon^2
\qquad
y_{11} \sim \epsilon^3
\qquad
y_{12} \sim \epsilon^2
\qquad
y_{21} \sim \epsilon
\qquad
\delta^{(\prime)} \sim \epsilon^2
\qquad
\gamma^{(\prime)} \sim \epsilon.
\end{align}
Note that $\epsilon$ is merely a fiducial quantity, not an expansion parameter of the model. We now apply the rotation (\ref{eq:misalignment:rotation}) and study the order of magnitude of the off-diagonal terms. By construction the terms proportional to $b$ are completely diagonalized. We consider the terms proportional to $c$ ($f_{b_L}$) and $d$ ($f_{b_R}$) separately.

\subsection[Misalignment from $f_{b_L}$]{\boldmath Misalignment from $f_{b_L}$}

First consider the terms proportional to $c$, which are split by the relative size of $f_{b_L}$ versus $f_{s_L}$ from internal zero mode propagators. 
The part of the $C_{7b}'^\dag$ term proportional to $c$ goes like
\begin{align}
\left.C_{7b}'^\dag\right|_c \sim 
  \left(y_{21} +  \gamma' \,y_{22}\right) - \delta' \left(\gamma'\, y_{12} + y_{11}\right).\label{eq:misalignment:c:7bp:c}
\end{align}
Naively the first term is of $\mathcal O(\epsilon)$ and appears to dominate the expression. This, however, does not account for relations coming from alignment.
Observe that the minus sign here comes from the choice of parameterization in  (\ref{eq:misalignment:estimate}). Further, observe that changing the relative sign in (\ref{eq:misalignment:c:7bp:c}) is equivalent to changing the sign of $c$ in the top row of (\ref{eq:misalignment:estimate}). In this case, however, the $c$ matrix would be completely aligned with the SM mass matrix and the off diagonal term (\ref{eq:misalignment:c:7bp:c}) would vanish. 
Thus the first and second terms in (\ref{eq:misalignment:c:7bp:c}) must be of the same order of magnitude in order for them to cancel when the relative sign is swapped---in other words, $(y_{21} +\gamma'\, y_{22}) \sim \epsilon^5$ in order to match the naive order of magnitude of the second term. We thus have
\begin{align}
c\left.C_{7b}'\right|_c \sim \epsilon^6.
\label{eq:misalignment:C7bp:c}
\end{align}
This observation reflects the key cancelation that causes the relative size of the primed and unprimed misalignment terms to differ from that of the anarchic terms of the amplitude.

The contribution to the $C_{7b}$ term proportional to $c$ is
\begin{align}
\left.C_{7b}\right|_c \sim 
	 \delta\left( \gamma \, y_{21}  +  y_{22}\right) -\left( \gamma \, y_{11} + y_{12}\right).\label{eq:misalignment:c:7b:c}
\end{align}
Unlike $C_{7b}'$, both terms in the above expression are dominated by their $\mathcal O(\epsilon^2)$ components and we find
\begin{align}
c\left.C_{7b}\right|_c \sim \epsilon\, \epsilon^2 = \epsilon^3,\label{eq:misalignment:C7b:c:final}
\end{align}
as expected from a naive estimate.

\subsection[Misalignment from $f_{b_R}$]{\boldmath Misalignment from $f_{b_R}$}

We perform the same analysis on the terms proportional to $d$, which implicitly encode the split between terms that carry factors of $f_{b_R}$ versus $f_{s_R}$ from internal propagators. For $C_{7b}$ we have 
\begin{align}
\left.C_{7b}\right|_d \sim 
	\left( y_{12} + \delta y_{22} \right) - \gamma \left( y_{11} + \delta\, y_{21}\right).\label{eq:misalignment:c:7b:d}
\end{align}
Following the argument that the terms should cancel when the sign is swapped and using this to estimate the size of each bracketed term, one finds $d\left.C_{7b}\right|_d \sim \epsilon^6$, so that the net contribution of the $d$ term is subdominant to (\ref{eq:misalignment:C7b:c:final}).

On the other hand, the $f_{b_R}$ misalignment in the $C_{7b}'$ term cannot be neglected,
\begin{align}
\left.C_{7b}'\right|_d \sim 
	\gamma'\left(\delta'\, y_{12} +  y_{22} \right) -  \left(\delta'\, y_{11} +  y_{21}\right).\label{eq:misalignment:c:7pb:d}
\end{align}
Here both terms are $\mathcal O(\epsilon)$ so that the total contribution is
\begin{align}
d \left.C_{7b}'^\dag\right|_d \sim \epsilon^3,
\end{align}
which dominates over the term proportional to $c$ in (\ref{eq:misalignment:C7bp:c}). 

\subsection{Size of misalignment coefficients}

Thus the final order of magnitude estimate for the $C_{7b}$ and $C_{7b}'$ coefficients are
\begin{align}
C_{7b} &\sim c \left.C_{7b}\right|_c \sim \epsilon^3\\
C_{7b}' &\sim d \left.C_{7b}'^\dag\right|_d \sim \epsilon^3,
\end{align}
so that unlike the anarchic contribution, the right-to-left (unprimed) and left-to-right (primed) Wilson coefficients are of the same order of magnitude.

\section{Comments on 5D dipole theory uncertainties}
\label{sec:theory:uncertainty}

Finite 5D loop effects carry subtleties associated with cutoffs and UV sensitivity\footnote{We thank K.~Agashe, J.~Hubisz, and G.~Perez for discussions on these subtleties.}. While the one loop contribution discussed in this paper is manifestly finite, higher loops are potentially divergent and require explicit calculations. Here we focus on the sensitivity of the finite loop-level result to UV physics at, for example, the strong coupling scale where the 5D theory is expected to break down. 
In \cite{Agashe:2006iy} it was pointed out that the naive dimensional analysis (NDA) for a brane and a bulk Higgs differ due to the dimension of the Yukawa coupling---the NDA two-loop contribution for the former gives an $\mathcal O(1)$ correction relative to the one loop result, whereas this is not expected for the latter. In this appendix we comment on subtleties coming from 5D Lorentz invariance that may plausibly avoid this `worst case' NDA estimate. 
Indeed, the NDA for the one-loop contribution to these dipole operators is logarithmically divergent; one may understand the correct one-loop finiteness as coming from 5D Lorentz symmetry. 

These comments are meant to demonstrate non-trivial points in these calculations that require particular care when drawing conclusions about UV sensitivity in these processes; a more careful investigation with explicit calculations of these effects is beyond the scope of this work.

Note that the general features of the phenomenological picture presented in Section~\ref{sec:num} are unchanged even if there are $\mathcal O(1)$ corrections to the Wilson coefficients.

\subsection{KK decomposition}

5D Lorentz invariance imposes that in the KK reduced theory, the 4D loop momentum cutoff should be matched to the number of KK modes in the effective theory. This was mentioned in \cite{Csaki:2010aj} to motivate a manifestly 5D calculation by pointing out that naively taking the finite 4D loop cutoff to infinity drops terms of the form $(n M_{\text{KK}}/\Lambda)^2$, where $nM_\text{KK}$ is approximately the mass of the $n^\text{th}$ KK mode. Indeed, from the 4D perspective this may appear to suggest a non-decoupling effect where the dominant contribution comes from heavy KK states so that the calculation seems to be sensitive to UV physics.

\begin{figure}[h!]
\begin{center}
\includegraphics[width=.4\textwidth]{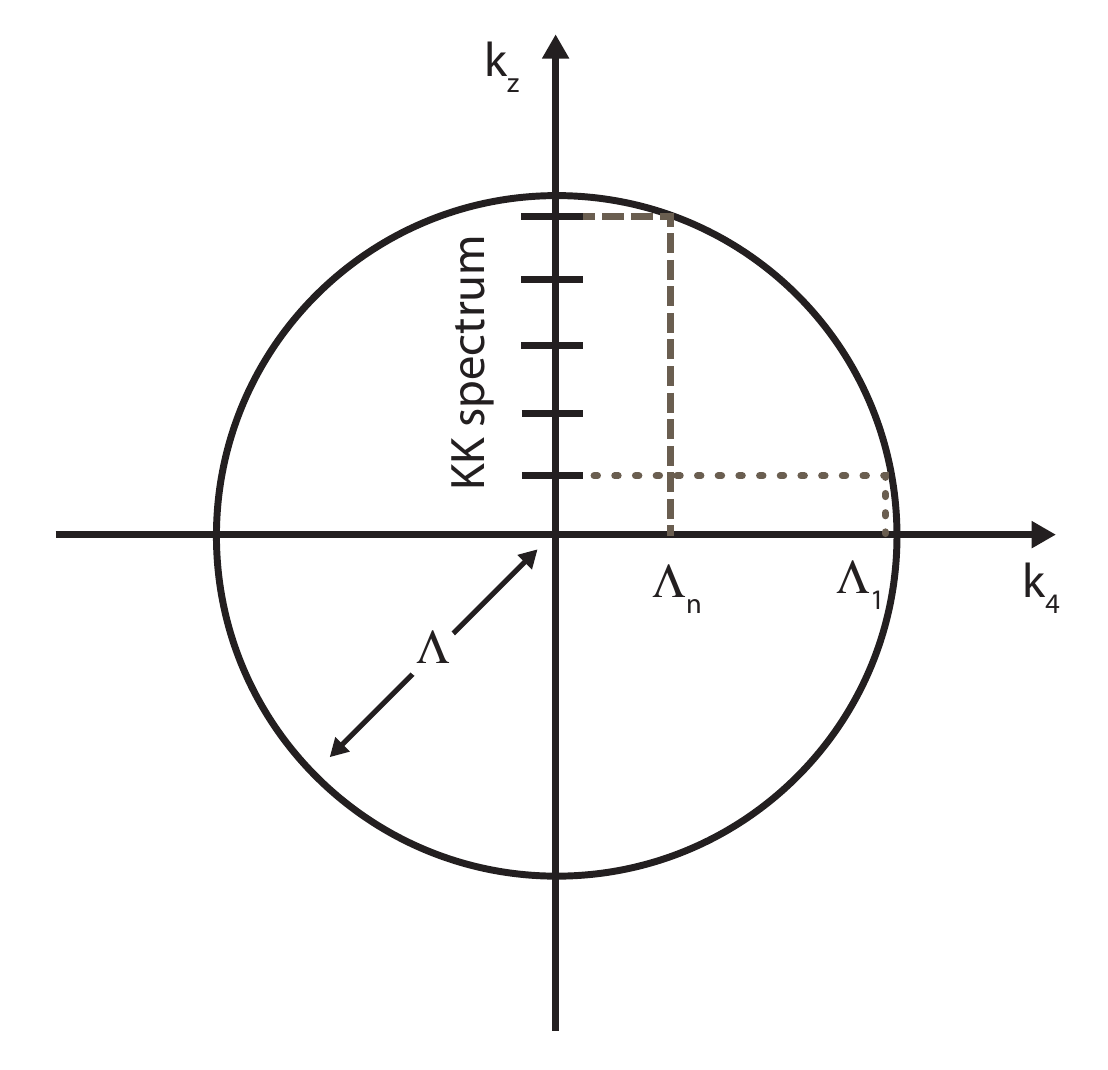}
\end{center}
\caption{A sketch of the 5D momentum space where the circle of radius $\Lambda$ represents the boundary of a 5D Lorentz invariant loop momentum integration region. Marks on the $k_z$ axis show the masses of KK states. Dashed lines demonstrate that the 4D loop cutoff which respects 5D Lorentz invariance depends on the particular KK mode.  \label{fig:5D:cutoff}}
\end{figure}

However, as demonstrated in Fig.~\ref{fig:5D:cutoff}, imposing 5D Lorentz invariance requires that each KK mode carries a different 4D momentum cutoff. 
In particular, the $n^\text{th}$ KK mode carries a smaller 4D cutoff $\Lambda_n$ than that of the first KK mode, $\Lambda_1$ since the momentum integral must fall within the circle of radius $\Lambda$, the 5D momentum space cutoff. 
Thus in 4D the high KK modes are not sensitive to the same cutoff as lower KK modes. This gives a sense in which 4D decoupling can manifest itself while preserving 5D Lorentz invariance.
In this sense it is difficult to use this matching to diagnose UV sensitivity.

As a qualitative and demonstrative estimate, one can use the expression in Section~6.6 of \cite{Csaki:2010aj} for a neutral Higgs diagram and impose a KK number dependent cutoff for each state in the loop so that 5D Lorentz invariance is imposed as in Fig.~\ref{fig:5D:cutoff}. One finds that, for example, in a sum of 200 KK modes, the highest 20 modes only contribute $\sim 20\%$ to the total result.

\subsection{5D cutoff}

\begin{figure}[h!]
\begin{center}
\includegraphics[width=.4\textwidth]{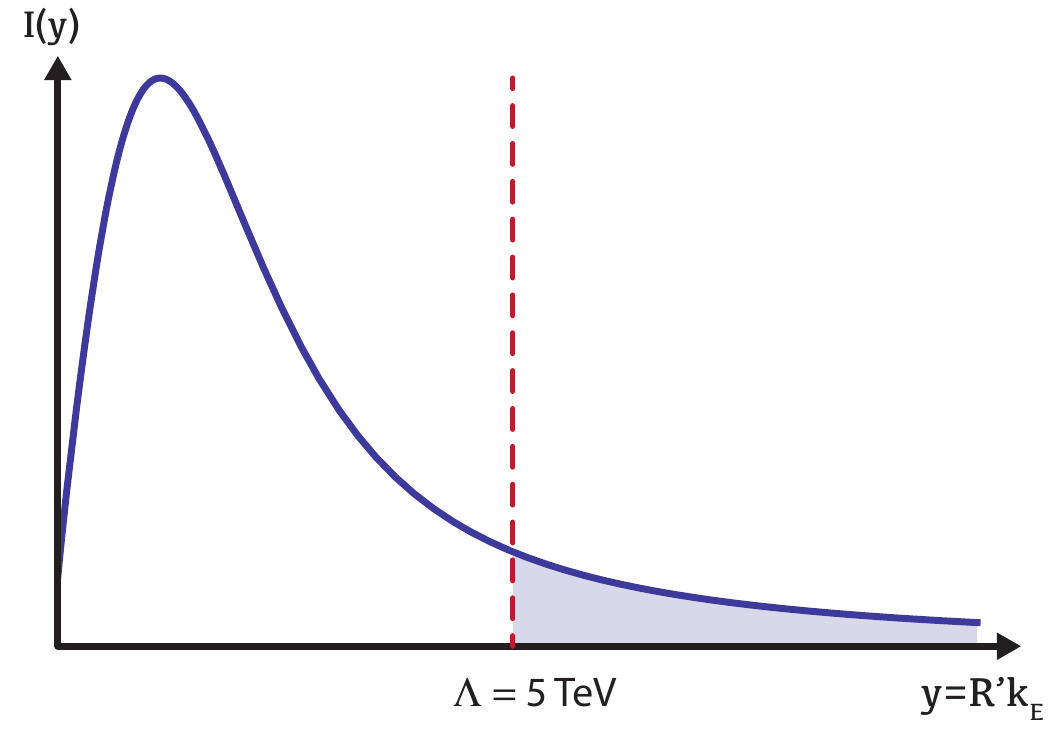}
\end{center}
\caption{Plot of the charged Higgs integrand as a function of the dimensionless loop momentum in the position/momentum space picture. The dashed line is a heuristic 5D cutoff $\Lambda$ representing the strong coupling scale. The shaded region represents the error from taking the loop momentum to infinity rather than $\Lambda$; the contribution of this shaded region is approximately $15\%$ of the total integral. \label{fig:5D:cutoffestimate}}
\end{figure}

Another way to diagnose UV sensitivity is to consider the effect of a cutoff in the 5D picture, for example, by setting a cutoff at $\Lambda = 5 \text{ TeV}$ representing the strong coupling scale at which the 5D theory breaks down. 
Fig.~\ref{fig:5D:cutoffestimate} shows the dimensionless integral associated with the charged Higgs loop, where $y=R' k_E$ is the dimensionless variable representing the loop momentum. Observe that the dominant contribution to the effect does not come from arbitrarily large $y$ but rather in the peak at low values of $y$. Cutting off the integral at $\Lambda = 5 \text{ TeV}$ (dashed line) gives an error of approximately $15\%$, which is comparable to the subleading diagrams that were not included in this analysis.

\FloatBarrier






\bibliographystyle{utphys.bst}
\bibliography{RSbsgamBib}


\end{document}